\documentclass[aps,prd,showpacs,nofootinbib,superscriptaddress,preprint,tightenlines,eqsecnum]{revtex4}
\usepackage{amsmath,amssymb,amsfonts}
\usepackage{accents}
\usepackage{graphicx}
\usepackage[usenames]{color}
\newcommand{\md}{{\mathrm d}}
\newcommand{\ed}{\md \!\!\!\! \md\, }

\newcommand{\be}{\begin{equation}}
\newcommand{\ee}{\end{equation}}
\newcommand{\tP}{\tilde{P}}

\begin{document}

\title{Canonical analysis of field theories in the presence of boundaries: Maxwell+Pontryagin}
\author{Alejandro Corichi}\email{corichi@matmor.unam.mx}
\affiliation{Centro de Ciencias Matem\'aticas, Universidad Nacional Aut\'onoma de
M\'exico, UNAM-Campus Morelia, A. Postal 61-3, Morelia, Michoac\'an 58090,
Mexico}

\author{Tatjana Vuka\v{s}inac}
\email{tatjana@umich.mx}
\affiliation{Centro de Ciencias Matem\'aticas, Universidad Nacional Aut\'onoma de
M\'exico, UNAM-Campus Morelia, A. Postal 61-3, Morelia, Michoac\'an 58090,
Mexico}
\affiliation{Facultad de Ingenier\'\i a Civil, Universidad Michoacana de San Nicol\'as de Hidalgo, 
Morelia, Michoac\'an 58000, Mexico}

\begin{abstract}
We study the canonical Hamiltonian analysis of gauge theories in the presence of boundaries. 
While the implementation of Dirac's program in the presence of boundaries, as put forward
by  Regge and Teitelboim, is not new, there are some instances in which this formalism is incomplete. Here
we propose an extension to the Dirac formalism --together with the Regge-Teitelboim strategy,-- that 
includes generic cases of field theories. We see that there are two possible scenarios, one where there is 
no contribution from the boundary to the symplectic structure and  the other case in which there {\it is} 
one, depending on the dynamical details of the starting action principle.
 As a concrete system that exemplifies both cases, we consider a theory that can be seen both as 
defined on a four dimensional spacetime region with boundaries --the bulk theory--, or as a theory defined 
both on the bulk {\it and} the boundary of the region --the mixed theory--. The bulk theory is given by 
the 4-dimensional  Maxwell + $U(1)$ Pontryagin action while the mixed one is defined by the 
4-dimensional Maxwell + 3-dimensional $U(1)$ Chern-Simons action on the boundary. Finally, 
we show how these two descriptions of the same system 
are connected through a canonical transformation that provides a third description. The focus here is in 
defining a consistent formulation of all three descriptions, for which we rely on the geometric 
formulation of constrained systems, together with the extension of the Dirac-Regge-Teitelboim (DRT) 
formalism put forward in the manuscript.  
\end{abstract}

\maketitle

\section{Introduction}
\label{Sec:1}

%Issue of a gauge field theory on a region with boundaries
%$\bullet$ Why bother about boundaries? 
In this manuscript we consider field theories, and gauge theories in particular, where the spacetime region  under consideration possesses a boundary. While most of the literature focuses on the case without boundaries, or they are neglected, theories with a boundary have also been the subject
of study in multiple contributions \cite{Regge&Teitelboim,Troessaert,BCE,ABOL,BBGS1,FPR,GS,HM,BBTS,Zabzine,Sheikh,Vergara}.  One may wonder why should one bother considering boundaries in the formalism. After all, one does not expect boundaries to contribute to the local dynamics in the inside, bulk, region. What one can learn from the extensive literature on the subject is that boundaries become very important, providing relevant physical information, and complementing in special ways the  information on the bulk.
%$\bullet$ {\color{blue} What about boundary degrees of freedom?}
One of the first issues one should consider when there are boundaries present is that of possible boundary degrees of freedom. Even when fields on the boundary are usually induced by those at the bulk, there
are instances in which {\it new} boundary degrees of freedom emerge. Furthermore, there are system where extra degrees of freedom are introduced to make the formalism well defined, or to force the system to
satisfy certain predetermined notions of symmetry. In any case, there are multiple examples of systems that posses degrees of freedom at the boundary.

%Is the Dirac method complete? Does it need to be extended?
%$\bullet$ {\color{red} Do we have to modify the Dirac-Regge-Teitelboim prescription?}
One important example is represented by diffeomorphism invariant theories, that is, theories that need no background structure in its standard Lagrangian formulation. It is a well known result that in such theories, within a covariant treatment, physical observables are represented by integrals over the boundary (for a discussion see \cite{CRV2}). Thus, it becomes particularly important to understand how to consistently define and treat gauge theories in regions with boundaries.
In this paper we shall focus our attention to Hamiltonian {\it canonical} methods for dealing with field theories. The standard method for treating field theories that are constrained was introduced by Dirac 
\cite{Dirac} and others \cite{others} in the middle of the XX century. Notably, those treatments did not consider boundaries. The standard extension of Dirac's formalism was put forward by Regge and Teitelboim \cite{Regge&Teitelboim} who were studying asymptotically flat general relativity in the canonical ADM formalism. This addition to the Dirac theory is sometimes regarded as standard in the treatment of gauge field theories with boundaries (see for instance the review article \cite{Troessaert} and references therein). However, there have been several papers suggesting that one needs to extend the Dirac formalism beyond that of the Regge-Teitelboim (RT) formulation, by defining extra structure at the boundary. For instance, there is an ample literature on the definition of boundary Poisson brackets \cite{MPB}, but those proposals seem to be, in many cases, somewhat ad-hoc, and have not been introduced into the mainstream of the literature. More relevant are certain ideas in
early treatments of isolated horizons in the canonical setting \cite{IH-old}, and particularly a series of recent papers by Barbero {\it et al} \cite{Barbero,Barbero1,Barbero2,JuanM}, who have extended the geometrical approach to constrained systems by Gotay, Nester and Hinds \cite{GNH} to the case of boundaries. Barbero and collaborators have pointed out, in some instances, the need to define certain geometrical quantities on the boundary that are not present in the standard RT formalism. Thus, there seems to be enough evidence for the need to extend the Regge-Teitelboim formalism for generic gauge theories with boundaries.
Despite this important progress, in our opinion there does not seem to be a consistent extension of the RT formalism, that is generic enough and that can reduce to the standard formalism for the cases where it is known to be successful. The main aim of this contribution is to put forward such an extension. 

%$\bullet$ When do we have a boundary contribution to the Symplectic Structure? 
In the standard Dirac-Regge-Teitelboim formalism, there is no boundary contribution to the symplectic structure, and the Hamiltonian formalism is made consistent by asking that the gradients of the relevant functionals (or variations in the standard lore) be well defined by having no contributions from the boundaries. The pressing questions are then: When will the theory have boundary contributions to the symplectic structure? How is that going to change the Regge-Teitelboim criteria? How does that affect the dynamics of the theory? Are there new boundary degrees of freedom? The first purpose of this manuscript is to put forward answers to these relevant questions. To be precise, in answering the first question above, we shall see that the relevant object to consider is the action, not surprisingly, and from that one can find in a natural way whether the theory acquires boundary terms in the symplectic structure. In a canonical setting, the pertinent action is of course, the {\it canonical action} defined on phase space.
Asking that the action and its variations be well defined guaranties that its corresponding Euler-Lagrange equations, that are nothing but Hamilton's equations, are well defined. In the symplectic language that we shall adopt here, this means that a key equation relating a special vector field and the gradient of a function should be satisfied. If there is a boundary contribution to the symplectic structure this implies having bulk and boundary equations of the motion. As it can already be guessed, all these requirements at the boundary already imply that the Regge-Teitelboim criteria has been transcended.

What we find is that, when there is a boundary, one can encounter two possible scenarios. The first one in which there is no boundary contribution to the symplectic structure, corresponds precisely to the well known Regge-Teitelboim case. As we shall see, consistency of the formalism is then reduced to their well know criteria, namely asking that variations of functionals have no boundary contribution. In the second case, the theory {\it will} have a boundary contribution to the symplectic structure, in which case the criteria for consistency involves possible boundary terms from the variations; differentiability does not equal vanishing boundary terms. As we shall see in the concrete example to be analysed, there is a {\it third} possible scenario. One could, for instance, start with a theory that has no boundary contribution to the symplectic structure and through a canonical transformation, arrive to an equivalent description in which one has ``acquired" a boundary term to the symplectic structure. In this new description one has to adopt the rules of the second case above, even when the original description had to be treated by the RT rules.

In our description of what the main issues under consideration are, we have put special attention to the 
symplectic structure $\Omega$ of the theory. This already hints that we have adopted a geometrical  approach to the problem, where the relevant space
describing the physical system is the canonical phase space endowed with a symplectic structure.  The standard object used in Hamiltonian dynamics, namely the Poisson bracket, becomes secondary and can be derived from the symplectic structure $\Omega$. As has been emphasized by Barbero {\it et al}, the use of geometrical methods for field theories with boundaries is particularly convenient. We have adopted this viewpoint that allows us to extend the DRT formalism in a systematic and straightforward way.

%We can learn from the geometical method and offer an extension

As an illustration of the general theory put forward here, we consider a system that can have two alternative description. As a bulk theory in 4 spacetime dimensions it is defined by the Maxwell + Pontryagin theory. Since the Pontryagin theory can alternatively be described by the Chern-Simons action on the 3-dimensional boundary, then the alternative description is given by the Maxwell + Chern-Simons theory. This second description of the system represents one of those cases in which the boundary contribution to the action possesses time derivatives of the basic variables, and thus the theory acquires a boundary contribution to the momenta and corresponding symplectic structure. What we shall see is that, even when both descriptions are quite different, in terms of variables and constraint structure, the boundary conditions for the phase space variables that need to be satisfied for the consistency of both descriptions coincide. This, together with the bulk equations of the motion (and bulk constraints) that are the same, warrant the same physical system regardless of the starting point. 
It is a well known fact that, in the case when there are no boundaries, one can ``absorb" the Pontryagin term via a canonical transformation and all that remains is the Maxwell theory. If the spacetime region has boundaries  this is not the case. What we find is that we can indeed start with the Maxwell + Pontryagin theory, and through a canonical transformation go to a Hamiltonian description  where the Maxwell contribution is isolated. Furthermore, we obtain  {\it a boundary contribution} to the symplectic structure.  This fact  forces one to adopt the ``second description" of the system and therefore to allow for boundary contributions to gradients and Hamiltonian vector fields, even when the Hamiltonian has only contributions form the bulk. In this way we recover the Maxwell + Chern-Simons theory.
What one learns is that even when one starts with a pure bulk theory, the canonical transformation introduces boundary terms and with them, triggers the extended formalism that we are introducing in this paper.
%$\bullet$ Is there a relation between boundary conditions and constraints?

%We can learn from the geometical method and offer an extension
The structure of the paper is as follows. In Sec.~\ref{Sec:2} we review the standard canonical Hamiltonian formalism from a geometric perspective, including field theories, and propose its extension for the case when boundaries are present. We also include a description for treating systems with constraints. In Sec.~\ref{Sec:3} we analyse the Maxwell + Pontryagin theory from the perspective of our formalism developed in Sec.~\ref{Sec:2}. The Maxwell + Chern-Simons theory is the subject of Section \ref{Sec:4}, while the canonical transformation that connects both descriptions is introduced in Sec.~\ref{Sec:5}. We end with a discussion in Sec.~\ref{Sec:6}. We present some useful formulae in an Appendix. 

In order to highlight the conceptual issues, we have considered without any loss of generality the simpler Abelian $U(1)$ case. The extension of our analysis to non-Abelian Yang-Mills theory is straightforward, and yields similar results. Throughout the manuscript we employ Penrose's abstract index notation and have considered fully covariant expressions for the action functionals and 3+1 decomposition, thus making our results valid on any (globally hyperbolic) spacetime and for arbitrary foliations and time evolution.

\section{Canonical Hamiltonian Formalism, Boundaries and Constraints}
\label{Sec:2}

%review of basic symplectic stuff
In this section we shall present a brief review of the canonical Hamiltonian treatment of constrained systems using the language of symplectic geometry. This section has four parts. In the first one we describe
the general framework of Hamiltonian dynamics introducing the relevant geometrical quantities. In the second one we focus on the field theory case, where one has to deal with the infinite dimensionality of the phase space. In the third one we introduce boundaries and describe how one has to adapt the Regge-Teitelboim prescription to accommodate for the most general case of field theories with boundaries. We end by including constraints and commenting on how to extend the Dirac formalism to the case where boundaries are present.
%Description of what taking functional derivatives, symplectic structure and Poisson Brackets mean

\subsection{Canonical Hamiltonian Formalism}
\label{Sec:2.A}

In this part we shall recall how one constructs the standard canonical Hamiltonian formalism using some geometrical ideas (for introductions to the symplectic formalism see, for instance, \cite{Arnold,Marsden}). The starting point for this procedure is
a configuration space ${\cal C}$, which is the space where the configuration variables ``live". 
For the approach to the subject we shall follow, there is a
fundamental object that we shall call the {\it  the momentum function}: $P:V^a\in T_{Q}{\cal C}
\mapsto \mathbb{R}$, where $T_Q{\cal C}$ is the tangent space to the configuration space ${\cal C}$ at the point $Q\in{\cal C}$. That is, if we extend its action to all of ${\cal C}$, then $P$ becomes a mapping from vector fields on configuration space to a function on configuration space, or in other word a (differential) 1-form on ${\cal C}$.
If we employ the abstract index notation for geometrical objects where $V^\alpha$ denotes a vector and $W_\alpha$ a 1-form, this map can be written as:
$P[V]:=P_{\alpha}\,V^{\alpha}$. 

Even when one starts from a configuration space ${\cal C}$, the relevant space where physical states are defined in Hamiltonian dynamics is a phase space $\Gamma$. The standard choice for the phase space is
to define it as:
\be
\Gamma:=T^*{\cal C},
\ee
that is, the cotangent bundle over ${\cal C}$.
Thus, the phase space $\Gamma$ consists of pairs $(Q,P)$, where $Q\in {\cal C}$ and $P$ is a 1-form on ${\cal C}$. In order to have the full structure needed for the Hamiltonian dynamics, we need to endow  $\Gamma$ with a symplectic structure $\Omega$. To do that, we
define a 1-form $\Theta$ on $\Gamma$ with the property that, when acting on the velocity vector $\dot{X}$, which is the tangent vector to the dynamical trajectories on phase space, the result coincides with the
action of the momentum $P$ acting on the velocity vector $\dot{Q}$ on configuration space. Thus, we
have,
 \be 
 \Theta[\dot{X}]:=P[\dot{Q}]\, ,
\ee
where $\dot{X}$ is the velocity vector on $\Gamma$ and $\dot{Q}$ is the velocity vector on ${\cal C}$\footnote{We can think of the velocity vector $\dot{X}$ on $\Gamma$ as made up of pairs $(\dot{Q},\dot{P})$, such that $\dot{X}=\dot{Q}^a\frac{\partial}{\partial Q^a}+\dot{P}_a\frac{\partial}{\partial P_a}$.}.
The 1-form $\Theta$ is called the symplectic potential. If we adopt some coordinates $Q^a$ for the point $Q\in{\cal C}$, and correspondingly ``coordinates" $P_a$ for the ``point" $P$, it can be defined as\footnote{ We have chosen the most common expression for the potential. The form of $\Theta$ is not unique, since we could have taken, for instance, $\Theta= -Q^a\md P_a$.}:
\be
\Theta := P_a\,\md Q^a\, .
\ee
The {\it symplectic structure} $\Omega$ on $\Gamma$ is then defined simply as,
\be 
\Omega:=\md \Theta\, .
\ee
For the $(Q^a,P_a)$, {\it Darboux}, coordinates, this two-form on phase space becomes $\Omega=\md P_a\wedge \md Q^a$.
By its definition, the two-form is closed ($\md\Omega =0$) and non-degenerate\footnote{It is non-degenerate since we have assumed that the coordinates $Q^a$ are complete in the sense that $\{ \md Q^a\} $ span the cotangent space of ${\cal C}$.}, so it can be taken as the canonical symplectic two-form.

Note that the definition of the symplectic potential and structure depends completely on the choice
of mapping $P$. How is one then to chose it? The answer comes from the canonical action, namely the action that defines the Hamiltonian theory, that can always be written in a form,
\be
S[Q,P]=\int \left[ P[\dot{Q}] - H(Q,P)\right] \md t\, ,
\ee
from which one can identify the mapping $P$. Here $H$ is the Hamiltonian of the theory, that is, the function on phase space responsible for dynamics, as we shall see below. Here we are not particularly interested on how one arrives to the canonical action but the standard procedure is to start from a Lagrangian action and perform a decomposition. In the examples that we shall consider, we will show that in detail.

Once we have endowed the phase space with a symplectic structure we can define relevant geometrical objects. The first one is a set of preferred vector fields $X$, that are called {\it Hamiltonian vector fields} (HVF). They have the property that they leave the symplectic structure invariant, in the sense that $X$ is Hamiltonian iff
\begin{equation}
\pounds_X\Omega =0\, .
\end{equation}
Now, using Cartan's identity $\pounds = i\md + \md i$, we have
%\be
%\pounds_X\Omega =(\md\Omega)(X,\cdot,\cdot) + (\md (\Omega(X,\cdot)))(\cdot) =\md (\Omega(X,\cdot))=0\, .
%\ee
\be
\pounds_X\Omega = i_X(\md\Omega )+ \md (i_X\Omega )=0\ \ \ \Rightarrow\ \  \ \md (\Omega (X,\cdot ))=0\, .
\ee
From this last equation it follows that, locally, there exists a function $F$, such that, 
\be 
\md F=\Omega(\cdot,X_F)\, ,\label{ecuacionimportante}
\ee
where we now denote by $X_F$ the Hamiltonian vector field associated to the function $F$. That is, there is a (-n almost, except for topological obstructions,) one to one correspondence between Hamiltonian vector fields and differentiable functions on $\Gamma$. In maths terms, one says that $X_F$ represents the symplectomorphism (or canonical transformation, in physics parlance), generated by the function $F$. 

Furthermore, physical observables are described by functions on $\Gamma$.
Thus, any observable has an associated vector field and therefore, generates a canonical transformation on $\Gamma$. From now on, we only consider Hamiltonian vector fields.
If we now contract with an arbitrary (Hamiltonian) vector field $Y$, we have,
\be
\pounds_Y F := \md F(Y)=\Omega(Y,X_F)\, .\label{lie-derivative}
\ee
If $Y$ itself is the HVF associated to the function $G$,
then we can define a product on the space of functions, the {\it Poisson Bracket} between $F$ and $G$ as,
\be
\{F,G\} := - \Omega(X_F,Y_G)\, .\label{PB}
\ee 
Dynamical evolution of the system, that is described by Hamilton's equations are then generated by a preferred function, namely the Hamiltonian $H$, through its associated HVF $X_H$,  and read,
\be
\dot{F} :=\pounds_{X_H} F=\{F,H\} := \Omega(X_H,X_F)\, ,
\ee
for any observable $F$ on $\Gamma$. Note that the HVF associated to the preferred function $H$ is what we were calling the velocity on $\Gamma$: $X_H=\dot{X}$.

\subsection{Field Theories}
\label{Sec:2.B}

Let us now consider the case of field theories. It is clear that the geometric ideas are the same, just that the dimensionality of the phase space is infinite. 
One has to be careful with this, but the spirit of the formalism is the same. Here we shall not focus on the functional analytic formulation of the problem, and shall assume that these issues can been taken care of (for example, as in [21]).

We shall take the  phase space  to have coordinates $(\phi,\tP)$. Here, for simplicity and without loss of generality, we are denoting by $\phi$ the configuration fields and by $\tP$ the corresponding momenta, but we are not restricting ourselves to a scalar field theory.
Now, if we assume there is no boundary,  the momentum mapping looks like
\be
P[V]=\int_\Sigma \md^3\! x \, \tilde{P}\,V\, .\label{moment-map}
\ee
where $V$ is, as before, a tangent vector to the configuration space, and $\Sigma$ is a spatial hypersurface in spacetime. The `tilde' over the symbol
$P$ indicates that it is a density of weight one. This means that the symplectic potential $\Theta$ is of the form,
\be
\Theta=\int_\Sigma\md^3x\,\tilde{P}\;\ed\phi\, ,\label{symplectic-potential-field}
\ee
where the symbol $\ed$ denotes the exterior derivative on $\Gamma$.  The symplectic structure can be written as,
\be
\Omega=\ed\Theta=\int_\Sigma\md^3x\,\,\ed \tilde{P}\wedge \ed\phi\, .
\ee
It is illustrative to see what the Hamiltonian vector fields look like in this case.
Let us start by considering a simple configuration function 
$\Phi=\phi[f]:=\int_\Sigma\md^3x\,f\,\phi$. Its gradient is then given by,
\be
\ed \Phi=\int_\Sigma\md^3x\,f\,\ed\phi\, .\label{gradientfunctionF}
\ee
A generic HVF can be written in the coordinate basis, as
\be
X= \int_\Sigma\md^3x\,\left[X^{\phi}(x)\left(\frac{\delta}{\delta\phi(x)}\right)+
X^{\tP}(x)\left(\frac{\delta}{\delta \tP(x)}\right)\right]\, ,\label{genericvector}
\ee
where $X^{\phi}(x)$ and $X^{\tP}(x)$ are the components of the vector field in the `directions' given by coordinates $\phi(x)$ and $\tP (x)$ respectively. Let us now compute the directional derivative of the function $\Phi$ as given above,
\begin{eqnarray}
X(\phi[f]):=\ed \Phi(X) &=& \int_\Sigma \int_\Sigma\md^3x\,\md^3y\,f(y)\,\ed\phi(y)\left[X^{\phi}(x)\left(\frac{\delta}{\delta\phi(x)}\right)+
X^{\tP}(x)\left(\frac{\delta}{\delta \tP(x)}\right)\right] \, \nonumber\\ \nonumber
& = & \int_\Sigma \int_\Sigma \md^3x\,\md^3y\,f(y)\,X^\phi(x)\,\delta^3(x,y)\,\\ 
&=& \int_\Sigma \md^3x\,f(x)\,X^\phi(x)\, ,\label{actionofX}
\end{eqnarray}
where we have used the equation $\ed\phi(y)\left(\frac{\delta}{\delta\phi(x)}\right)=\delta^3(x,y)$.
The last equality of (\ref{actionofX}) is sometimes written as,
\be
\delta \Phi=\int_\Sigma\md^3x\,f(x)\,\delta\phi(x)\, ,
\ee
but the correct interpretation is that of Eq.(\ref{actionofX}), namely as a directional derivative. Similarly, we can consider the linear momentum function, already defined, $P[V]=\int_\Sigma \md^3x \tilde{P}\,V$, and see that
\be 
X(P[V])=\int_\Sigma \md^3x\,V(x)\,X^{\tP}(x) \,  .
\ee
Or, in the standard notation found in the literature, $\delta P[V]=\int_\Sigma \md^3x\,V(x)\,\delta\tilde{P}(x)$.

It is also illustrative to compute the right hand side of Eq.(\ref{ecuacionimportante}) in the case of fields. We have,
\begin{eqnarray}
\Omega(\cdot,X) &=& \int_\Sigma \md^3x\; \ed \tP (x)\wedge\ed\phi(x)\left(\;\;\cdot\;\; ,
\int_\Sigma \md^3y\,X^\phi(y)
\frac{\delta}{\delta\phi(y)} + X^{\tP} (y)\frac{\delta}{\delta \tP (y)}\right)\, \nonumber\\
&=& \int_\Sigma\int_\Sigma \md^3x\, \md^3y\,\left( X^\phi(y)\delta^3(x,y)\, \ed \tP (x) -  X^{\tP}(y)\delta^3(x,y)\,\ed\phi(x)  \right) \, \nonumber\\
&=& \int_\Sigma \md^3x\,\left( X^\phi(x)\,\ed \tP (x) -  X^{\tP}(x)\,\ed\phi(x) \right)\, ,
\end{eqnarray}
from which we can now contract with another vector field $Y$ with ``components" $(Y^\phi(x),Y^{\tP}(x))$, and obtain
\be
\Omega(Y,X)= \int_\Sigma \md^3x\,\left(  X^\phi(x)\,Y^{\tP}(x) -   X^{\tP}(x)\,Y^\phi(x) \right)\, .
\ee

Let us now write down the form of the gradient $\ed F$ of a generic function $F$,
\be
\ed F=\int_\Sigma \md^3x\;\left(\frac{\delta F}{\delta\phi(x)}\,\ed\phi(x)+\frac{\delta F}{\delta \tP (x)}\,
\ed \tP (x)\right)\, ,
\ee
where $\frac{\delta F}{\delta\phi(x)}$ is the generalization to field space of the partial derivative,
that is, the directional derivative along coordinate directions. We are now in position to write down Eq.(\ref{ecuacionimportante}),
\begin{eqnarray}
\ed F &=& \Omega(\cdot, X_F)\, , \qquad {\rm as} \nonumber\\
\int_\Sigma \md^3x\;\left(\frac{\delta F}{\delta\phi(x)}\,\ed\phi(x)+\frac{\delta F}{\delta \tP (x)}\,
\ed \tP (x)\right) &=&  \int_\Sigma \md^3x\,\left( X_F^\phi(x)\,\ed \tP (x) - X_F^{\tP}(x)\,\ed\phi(x) \right)\, ,\nonumber\\
\end{eqnarray}
from which we can read-off the components of the Hamiltonian vector field $X_F$ associated to $F$ as:
\be
X_F^\phi(x) =  \frac{\delta F}{\delta \tP (x)}\, ,\label{comp-1}
\ee
and
\be
X_F^{\tP} (x) = - \frac{\delta F}{\delta \phi(x)}\, .\label{comp-2}
\ee
We can now replace (\ref{comp-1}) and (\ref{comp-2}) in the generic form of the vector field $X_F$,
which now takes the form
\be
X_F=\int_\Sigma \md^3x\;\left(\frac{\delta F}{\delta \tP (x)}\;\frac{\delta{}}{\delta\phi(x)}  - \frac{\delta F}{\delta\phi(x)}\;\frac{\delta}{\delta \tP (x)}  
 \right)\, .
\ee
Let us now act with $X_F$ on another function $G$, which yields
\be
\pounds_{X_F}G = X_F(G) = \int_\Sigma\md^3x\;\left( \frac{\delta F}{\delta \tP(x)}\;\frac{\delta{G}}{\delta\phi(x)} 
- \frac{\delta F}{\delta\phi(x)}\; \frac{\delta G}{\delta \tP(x)} \right)= \Omega(X_F,X_G)\, .\label{omega-contracted}
\ee
If we now compare Eq.(\ref{omega-contracted}) with (\ref{lie-derivative}) and (\ref{PB}) we can conclude that it corresponds to the Poisson bracket $\{ G,F\}$ as defined by Eq.(\ref{PB}). Thus, we have derived the standard form of the Poisson bracket as defined in the literature, but recall that we have just computed the right hand side of (\ref{PB}) which is the geometrical object that has an invariant geometric meaning. As mentioned above, we have assumed that $\Sigma$ has no boundaries and therefore, there are no ambiguities when computing the gradient of a function, since any potential boundary term is neglected and thus all expressions are as written (that is, as integrals over $\Sigma$). Let us now consider the case when boundaries are present.

\subsection{Boundaries}
\label{Sec:2.C}

%Why are boundaries relevant? Because in the standard Dirac analysis of gauge field theories, one normally
%disregards boundaries, and all the boundary terms that appear as one integrates by parts are discarded. 

In this part we shall consider the case when there is a boundary of the spatial hypersurface $\Sigma$. In this case one can no longer disregard the boundary terms that appear as one integrates by parts while computing the gradients of functions.
In the standard Regge-Teitelboim (RT) analysis of field theories with boundaries \cite{Regge&Teitelboim}, the main theme is to make all functio(al)s differentiable. This means that, when one computes the gradient $\ed F$ of the function $F$, there should be no contributions from the boundaries. This is equivalent to saying that all boundary terms that appear when taking the ``variation" of the phase space function, should vanish.
The unstated assumption is that this approach to field theories with boundaries is sufficient to deal with
all cases, namely that the RT method is generic. In this part we shall question this assumption and argue that there are cases in which one {\it has} to extend the formalism and allow for boundary terms to
``remain" throughout the canonical analysis. 

The first question one should ask is how one can tell when a theory will have contributions from the boundaries. Does it imply a boundary contribution to the symplectic structure? Non vanishing contributions to the gradient? In what follows we shall
try to answer those questions. As we shall argue, when there is a boundary present, there are instances in
which there shall be no contributions from the boundary, thus falling into de RT case, and some other cases in which there {\it will} be boundary contributions.

The key idea here is to recall the momentum map that we defined in the previous part. In Eq.(\ref{moment-map}) we had the expression with no contribution from the boundary. Let us now suppose that, in the canonical action we obtain, it contains a map that looks like,
\be\label{momentum-map-boundary}
P[V]=\int_\Sigma\md^3x\,\tilde{P}^aV_a + \int_{\partial\Sigma}\md^2x\, \tilde{P}_\partial^av_a\, ,
\ee
namely, we are assuming that, apart from the standard bulk term, there is as a contribution from the boundary $\partial\Sigma$. Now we are using coordinates
$(A_a,\tP^a)$ for the phase space (bulk) degrees of freedom, and  $(A^\partial_a,\tP^a_\partial)$ for the boundary degrees of freedom (Boundary DOF). 
As before, $V_a$ are the bulk components of an arbitrary tangent vector in configuration space, while $v_a$ represent its boundary components.
We have changed the notation, without any loss of generality, to bring it closer to the example we shall see in following sections.
This extra term implies  that the symplectic potential will acquire a boundary contribution 
\be
\Theta=\int_\Sigma\md^3x\,\,\tilde{P}^a\; \ed A_a + \int_{\partial\Sigma}\md^2x\,\, \tilde{P}_\partial^a\; \ed A^{\partial}_a\,
\label{symplectic-potential-boundary} .
\ee
From which the symplectic structure,
\be
\Omega=\ed \Theta= \int_\Sigma\md^3x\,\,\ed\tilde{P}^a\wedge\; \ed A_a + \int_{\partial\Sigma}\md^2x\,\, \ed\tilde{P}_\partial^a\wedge\; \ed A^{\partial}_a\, ,\label{omega-with-boundary}
\ee
also has a boundary term. Note that we are regarding the bulk and boundary degrees of freedom as independent, which will be reflected on our computation of the gradients, even when one (or more) of the boundary fields might have come from the pullback of bulk degrees of freedom. We shall encounter this situation in the example that we shall examine below.

%What are the practical implications of such a boundary contribution?}\\
Let us now explore what are the consequences of having such boundary terms in the symplectic structure.
Recall the basic Hamiltonian equation,
\be
\ed F(Y)=\Omega(Y, X_F)\, .\label{last-equation}
\ee
If there are no boundary terms in the right hand side of (\ref{last-equation}), then for the equations to be satisfied, there should be no boundary terms in the left hand side, that is, in the gradient. This is precisely the standard Regge-Teitelboim case, where we require the boundary terms in the gradient to vanish. In this case, all equations in the previous part are valid, even when there is a boundary present.

Let us now consider the case where we have boundary terms in the RHS of (\ref{last-equation}) due to the existence of boundary contributions to $\Omega$ as in (\ref{omega-with-boundary}). In order to have
(\ref{last-equation}) satisfied, we {\it must} have boundary terms in the LHS, namely, in the gradient.  The traditional sense of differentiability as advocated by Regge-Teitelboim is transcended and one has to admit a more general meaning to it. %(See \cite{Barbero1}). 
This also implies that one will have boundary contributions to the Hamiltonian vector fields such that, when contracted with $\Omega$, boundary terms {\it will} survive in general. This is one of the main observations of this note.

Let us now see how the formalism gets modified by these extra terms. Let us first consider a generic HVF
that takes the form.
\begin{eqnarray}
X &=& \int_\Sigma\md^3x\,\left[X^{A}_a(x)\left(\frac{\delta}{\delta A_a(x)}\right)+
X^{\tP a}(x)\left(\frac{\delta}{\delta \tP^a(x)}\right)\right] \nonumber\\
&+& \int_{\partial\Sigma}
\md^2y\,\left[ X^{A^\partial}_a(y)\left(\frac{\delta}{\delta A^{\partial}_a(y)}\right)+
X^{{\tP}_{\partial} a}(y)\left(\frac{\delta}{\delta {\tP}_{\partial}^a(y)}\right)\right]
\, .\label{genericvectorboundary}
\end{eqnarray}
The gradient of an elegible function $F$ will now have the form,
\begin{eqnarray}
\ed F &=& \int_\Sigma \md^3\!x\;\left(\frac{\delta F}{\delta A_a(x)}\,\ed A_a(x)+\frac{\delta F}{\delta \tP^a (x)}\,\ed \tP^a(x)\right)\nonumber \\
&{}& + \int_{\partial\Sigma}\md^2\!y\;\left(\frac{\delta F}{\delta A^{\partial}_a(y)}\,\ed A^{\partial}_a(y)
+\frac{\delta F}{\delta \tP^a_{\partial}(y)}\,\ed \tP^a_{\partial}(y)\right)\, , \label{gradconfrontera}
\end{eqnarray}
with its corresponding contribution from the boundary. As before, the quantities $\frac{\delta F}{\delta A^{\partial}_a(y)}$ should be though as a generalization of the partial derivative of the function $F$ along the coordinate $A^{\partial}$ and leaving the rest of the coordinates (including the bulk DOF) constant.
Note that we are assuming that the bulk and boundary degrees of freedom are {\it independent}, when taking directional derivatives, even when some of the boundary degrees of freedom might have arisen as the pullback of bulk DOF to the boundary. An important issue to note in the previous equation is that
the boundary contributions to the exterior derivative of $F$ might arise from both the direct derivation of a boundary term, {\it and } from a bulk term after integration by parts. Also, it should be noted that  Eq. (\ref{gradconfrontera}) restricts the possible boundary terms in the gradient to only those involving variables that are explicitly contained in the symplectic structure. For the rest of the variables, the standard RT criteria of requiring no boundary terms in the gradient also applies. Finally, note that Eq. (\ref{gradconfrontera}) can also be seen as defining what functionals on phase space are to be {\it differentiable}, in the sense that their exterior derivative is well defined. As mentioned now several times, we are encountering a scenario in which differentiability does not equal vanishing of boundary terms in the gradient.

Let us now contract the symplectic structure with a generic vector field,
\begin{eqnarray}
\Omega(\cdot,X) &=&
 \int_\Sigma \md^3\!x\,\left( X^{A}_a(x)\,\ed \tP^a(x) - X^{\tP a}(x)\,\ed A_a(x)\right)\nonumber\\
&{}& +
 \int_{\partial\Sigma} \md^2\!y\,\left( X^{A^\partial}_a(y)\,\ed \tP_\partial^a(y) - X^{\tP_{\partial} a}(y)\,\ed A^\partial_a(y)\right)\, .
\end{eqnarray}
By considering Eq.~(\ref{ecuacionimportante}) we can read off,  the bulk contribution of the HVF given by Eqs.~(\ref{comp-1}) and (\ref{comp-2}) that now take the form, 
\begin{equation}
(X_F)^{A}_a\!(x) = \frac{\delta F}{\delta \tP^a(x)}\;\;\;\; ; \;\;\;\; 
(X_F)^{\tP {a}}\!(x) = - \frac{\delta F}{\delta A_a(x)}\, .\label{comp-3}
\end{equation}
while the boundary components of the Hamiltonian vector field are given by,
\begin{equation}
(X_F)^{A^{\partial}}_a\!(y) = \frac{\delta F}{\delta \tP_\partial^a(y)}\;\;\;\; ; \;\;\;\; 
(X_F)^{\tP_{\partial} a}\!(y) = - \frac{\delta F}{\delta A^{\partial}_a(y)}\, .\label{comp-4}
\end{equation}
We are now in position of computing the totally contracted symplectic structure, namely, the Poisson Bracket between the functions $G$ and $F$,
\begin{eqnarray}
\{ G, F\} := \Omega(X_F,X_G) &=& \int_\Sigma\md^3\!x\;\left( \frac{\delta F}{\delta \tP^a(x)}\;\frac{\delta{G}}{\delta A_a(x)}  -  \frac{\delta F}{\delta A_a(x)}\;
\frac{\delta G}{\delta \tP^a(x)} \right)\nonumber\\
&{}& + \int_{\partial\Sigma}\md^2\!y\;\left( \frac{\delta F}{\delta \tP_{\partial}^a(y)}\;\frac{\delta{G}}{\delta A_a^{\partial}(y)} - \frac{\delta F}{\delta A_a^{\partial}(y)}\; \frac{\delta G}{\delta \tP_{\partial}^a(y)}
 \right)\, .
\end{eqnarray}
As we could have anticipated, the Poisson Bracket acquires a boundary term as a direct consequence of having a non-trivial contribution at the boundary in the momentum map and symplectic potential 
(\ref{symplectic-potential-boundary}). It should be stressed, however, that the previous expression was
derived from the momentum map, which in turn is obtained from the canonical action of the theory under
consideration. Furthermore, the boundary degrees of freedom are those present in the original action principle
and not external degrees of freedom. This has to be contrasted with several approaches 
that either postulate a surface Poisson Bracket or/and postulate new degrees of freedom at the boundary. 

Let us summarize the situation for a field theory defined on a region with boundaries. The starting point are always some bulk degrees of freedom $(A_a(x),\tP^a(x))$, and boundary degrees of freedom $(A^{\partial}_a(y),\tP_{\partial}^a(y))$. In certain instances some of the boundary degrees of freedom might just be the restriction/pullback of the bulk DOF to the boundary. We have two possibilities depending on the form of the action:\\
i) The momentum map (and thus the symplectic potential) only has a bulk contribution. In this case both the symplectic structure and all gradients must have vanishing contribution from the boundary. This corresponds precisely to the standard Regge-Teitelboim scenario;\\
ii) There is indeed a contribution from the boundary to the symplectic potential. In this case we acquire a boundary contribution to the symplectic structure, Hamiltonian vector fields, gradients and Poisson brackets. This situation leads us to regard the phase space $\Gamma$ as being a product of the {\it bulk} phase space $\Gamma_{\mathrm{Bulk}}$ and a {\it boundary} phase space $\Gamma_{\mathrm{Bound}}$, becoming of the form: $\Gamma=\Gamma_{\mathrm{Bulk}}\times\Gamma_{\mathrm{Bound}}$, where each part is to be thought of as independent from the other. 
Correspondingly, the symplectic structure (and PB) will be a sum of a bulk and a boundary term, each term ``acting" on their corresponding spaces: $\Omega=\Omega_{\rm Bulk} + \Omega_{\rm Bound}$.

%Regge-Teitelboim needs to be revisited/extended!
Let us end this discussion with an observation. One might wonder why one has abandoned, in the case 
ii)  we are considering above, the sound principle of asking the functions to be differentiable in the RT 
sense. While this is indeed true, upon reflection one realizes that the object we are requiring to be 
differentiable, for  Hamilton's equations (\ref{ecuacionimportante}) to be satisfied, is the {\it 
canonical action} given by $\int (P[\dot{Q}]-H) \md t$. Note that the boundary terms that survive under 
variations of the first term, cancel with the boundary terms remaining in the variation of the 
Hamiltonian, thus killing all boundary terms in the variation of the action.
Just as we needed those boundary terms to remain in Eq.~(\ref{ecuacionimportante}), we need 
them here as well. Thus, we recover full consistency of the formalism.

\subsection{Constraints}
\label{Sec:2.D}

Most physical theories are field theories where field configurations are constrained and can not have arbitrary values. Of particular relevance are gauge theories, where the fundamental variables are connections that have to satisfy certain relations and have a gauge symmetry, namely a symmetry that leaves the physics invariant. How to deal with these theories from the canonical perspective is the subject of Dirac's old analysis \cite{Dirac} (see also \cite{Pons} for a recent discussion). One important limitation of the formalism is that it only applies to the case where there are no boundaries. First steps in treating gauge theories in regions with boundary, for the case of general relativity with asymptotic boundary conditions, were taken in the very important contribution of Regge and Teitelboim \cite{Regge&Teitelboim}, that has already been mentioned. Other authors have applied the Regge-Teiteboim approach to other gauge theories
(see for instance the summary paper \cite{Troessaert} and references therein). However, a generic treatment of the subject when there is a contribution to the symplectic structure from the boundary, as discussed in Sec.~\ref{Sec:2.C} is, to the best of our knowledge, not available. The most complete treatment can be found in the work of Barbero {\it et al} \cite{Barbero,Barbero1,Barbero2} where they follow on the geometrical analysis of Gotay, Nester and Hinds (GNH) \cite{GNH}. The idea here is to recast Dirac's algorithm for constrained systems in the geometric language as described in previous sections. 
%while Barbero {\it et al} aim is to extend the GNH formalism to the case of regions with boundaries. There are, however some important differences between their approach and ours that we shall point out below.

Let us start by recalling the main elements of Dirac's procedure. We shall then describe the formalism in geometrical terms, and finally propose an extension to the case of boundaries. 
The canonical analysis follows the standard procedure of defining the Legendre transform of the Lagrangian, where as a first step the canonical momenta $P_a$ are defined as functions of $(Q^a,\dot{Q}^a)$. For regular systems, one can invert these relations and solve for the velocities $\dot{Q}^a$.
However, if the system is singular, this is no longer possible. In particular, if one of the momenta is of the form $P_k=\chi_k(Q)$, that is, it does not depend on the velocity, one has a constraint of the form $\phi_k=P_k - \chi_k(Q)=0$, which relates the canonical coordinates. All constraints that appear in this step are called {\it primary constraints}. From the geometrical perspective, one is stating that the physical configurations are not all points of $\Gamma$, but rather a subset that we shall denote by $\Gamma_0\subset \Gamma$, defined by the null set of the primary constraints. 

The first realization by Dirac was that one has to supplement the canonical Hamiltonian $H_{\rm c}$, as obtained by the Legendre transform, with a linear combination of the primary constraints $\phi_k$. The resulting total Hamiltonian, $H_T$, is of the form:
\be
H_T = H_c + u^k\phi_k \, ,
\ee
with the idea of implementing the constraints (a la Lagrange) in the canonical action. Of course, one needs to ensure that the dynamical evolution  preserves the constraints, or in other words, the corresponding Hamiltonian vector field $X_T$ should be tangent to $\Gamma_0$ and must satisfy,
\begin{equation}
\md H_T=\Omega(\cdot,X_T)\, .\label{ecuacionimportanteH}
\end{equation}
The vector field $X_T$ might not be tangent to $\Gamma_0$ (if it is, we are done and we only have the primary constraints), so we are faced with further possibilities. One of them is that we encounter new constraints $\phi_m=0$ that need to be satisfied. This means that we have to restrict ourselves to a submanifold $\Gamma_1\subset\Gamma_0$, where we expect the new Hamiltonian vector field (now associated to a new Hamiltonian $H_T + u^m\phi_m$) to be tangent and satisfy (\ref{ecuacionimportanteH}). We now need to ensure that these new constraints are preserved, so we have a process that is to be repeated until one does not encounter further constraints. Along the way, we might have found that the consistency conditions can be satisfied, provided one imposes  specific values for some of the multipliers $u^n$ (more on this below). Geometrically we are looking for the largest submanifold of $\Gamma$ for which there is a HVF $X_E$, associated to the {\it extended Hamiltonian} $H_E$, which is tangent to it, and that satisfies Eq.~(\ref{ecuacionimportanteH}) (for $H_E$ and $X_E$). 

Suppose that we have come to the end of the algorithm and we have found a set of constraints $\phi_n$, that define the constrained phase space $\bar{\Gamma}$. These constrains can be of two types, as classified by Dirac: they can be of the {\it first class} or {\it second class}. In general, a constrained system will have both types of constraints, so it is important to understand how one approaches each case. The standard procedure is to deal with the second class constraints first and then consider those of the first class. That is what we shall do here as well. 

A set of constraints $\phi_i$ is said to be of the  second class if the pullback of $\Omega$ to the corresponding submanifold defined by the vanishing of the
$\{ \phi_i \}$ constraints, is {\it non-degenerate}. Thus, we can restrict ourselves to the corresponding submanifold $\Gamma_D\subset\Gamma$, and consider the pair $(\Gamma_D,\Omega_D)$ as a new phase space where dynamics is unfolding. If all constraints were of the second class, then in the Dirac reduction algorithm we would have found that we could satisfy the consistency conditions by choosing some specific values for the multipliers $u^n$. The dynamics is fully determined and we have a Hamiltonian responsible for it. In other words, there is a unique $X_{H_D}$ that is tangent to $\Gamma_D$ and satisfies Eq.(\ref{ecuacionimportanteH}). Note that for the symplectic structure to be non-degenerate, we need an even number of constraints, so second class constraints always come in ``pairs".

Things become ``interesting" when the pullback of the symplectic structure to the submanifold defined by the constraints is degenerate. Suppose we have $l$ independent constraints $\phi_i$. We say they are of the first class if the submanifold $\Gamma_g$ defined by the vanishing of them is such that: The $l$ HVF $X_i$ associated to the constraints are tangent to $\Gamma_g$, and the pullback $\Omega_g$ of $\Omega$ to $\Gamma_g$ is degenerate with exactly $l$ degenerate directions. Recall that $X$ defines a degenerate direction of a pre-symplectic structure if $\Omega(Y,X)=0$, for all $Y$ tangent to $\Gamma_g$. What is the physical relevance of first class constraints? The answer is two fold. First, they signal the existence of {\it gauge} symmetries, and second, they imply that the dynamics is not uniquely determined by a vector field $X_E$. There are many allowed vector fields. Let us now see how these two consequences of first class constraints come about.

Let us start by understanding what gauge directions are. The first thing to notice is that the HVF associated to the $l$ constraints are not only tangent to $\Gamma_g$, but are also the degenerate directions of $\Omega_g$. Take for instance a constraint $\phi_i$ and its HVF $X_i$. Note that $X_i$ is tangent to the manifold $\Gamma_g$, which means that the directional derivatives along $X_i$ of all constraints $\phi_i$ should vanish. That is,
\be 
\pounds_{X_i}\phi_j=\Omega_g(X_i,X_j)=\{\phi_j,\phi_i\} \approx 0\, ,
\ee
where the symbol $\approx$ means identity on the constrained surface. This is the standard definition given by Dirac. Let us now see that the degeneracy of $\Omega_g$ along $X_i$ follows. Let us take an arbitrary tangent vector $Y$, and let us compute $\Omega_g(Y,X_i)$:
\be
\Omega_g(Y, X_i)= \pounds_Y \phi_i = 0\, ,
\ee
since all $\phi_i$ are constant on $\Gamma_g$. Let us now consider {\it physical} observables. An observable $f$ is said to be physical if it leaves the constrained surface $\Gamma_g$ invariant, namely if its associated vector field $X_f$ is tangent to $\Gamma_g$ (note that this is also true for the submanifold $\Gamma_D$ defined by second class constraints). Thus, since $X_i$ is a degenerate direction we have,
\be
0\approx \Omega_g(X_i,X_f)=\pounds_{X_i}f\, .
\ee
That is, any physical observable has to be constant along the integral curves generated by the vector fields $X_i$. Furthermore, these directions are known as {\it gauge directions} and the corresponding integral curves as {\it gauge orbits}. To summarize, the $l$ constraints $\phi_i$ generate canonical transformations that leave any physical observable invariant, and are therefore regarded as generators of gauge transformations. This discussion also tells us what is gauge and what not. Only transformations along degenerate directions of the pre-symplectic structure $\Omega_g$ are to be regarded as gauge. This trivial observations shall be relevant when one is considering field theories with and without boundaries. Note also that the gauge vector fields are integrable, so the gauge directions define actually a submanifold of $\Gamma_g$.

Let us now see how the dynamics is not uniquely determined in this case, and how that relates to the issue of gauge invariance. Let us assume that we have found a HVF $X_E$ with the properties we were seeking, namely tangent to the constraint surface $\Gamma_g$ and satisfying the equations
\be
\md H_E=\Omega(\cdot,X_E)\, ,
\ee
where $H_E$ is the  extended Hamiltonian. If we now add to $X_E$ any vector proportional to a gauge vector field (i.e. the HVF associated to a constraint) $\tilde{X}_E = X_E+\lambda X_i$, and compute the evolution equation of any physical observable $f$ with respect to $\tilde{X}_E$ we get,
\begin{eqnarray}
\pounds_{\tilde{X}_E} f= \Omega_g(\tilde{X}_E, X_f) &=& \Omega_g(X_E+\lambda X_i, X_f)=\Omega_g(X_E,X_f)+\lambda\Omega_g(X_i,X_f)\notag \\
&\approx& \Omega_g(X_E,X_f)=\pounds_{X_E} f\, .
\end{eqnarray}
That is, as far as physical observables are concerned, both $X_E$ and $\tilde{X}_E$ generate the same time evolution. Note that adding a vector proportional to a gauge vector field as we did, is the same as adding to the extended Hamiltonian $H_E$ a term proportional to the constraint $\phi_i$: $\tilde{H}_E=H_E+\lambda\phi_i$. Clearly, the same is true for any linear combination of first class constraints. It should be noted then that there is no such thing as a unique extended Hamiltonian $H_E=H_T+u^i\phi_i$, since the multipliers $u^i$ are {\it arbitrary}. For each choice of $\{u^i\}$ there is an extended Hamiltonian $H_E$ and a choice of HVF $X_E$. What we have seen is that {\it any} choice generates the same physical dynamics (as seen by physical observables).
Thus, we can conclude that dynamics is not uniquely determined in the sense that there is not a unique Hamiltonian, nor evolution vector field that generate dynamics, and that is because of the existence of gauge directions that are a consequence of having first class constraints in the first place.

\vskip0.3cm
\noindent{\it Boundaries}. Let us now describe how the existence of boundaries affect the Dirac-Regge-Teitelboim (DRT) formalism. As we have described in Sec.~\ref{Sec:2.C}, there are two possible scenarios when dealing with a field theory in a region with boundaries:\\
In case a) we {\it do  not} have contributions from the
boundary to the symplectic potential nor symplectic structure.\\ 
In case b) we {\it do} have a boundary contribution to both geometrical quantities.

The first step of the process is the definition of the primary constraints. If we are in case a), then we will only have primary constraints defined on the bulk, while in case b) there might be, apart from the primary constraints in the bulk, other primary constraints that pertain only to the boundary degrees of freedom as defined before.

The second step is to define the total Hamiltonian by adding a linear combination of the primary constraints to the canonical Hamiltonian, and to ask that all primary constraints be preserved during time evolution. We first have to construct the HVF $X_k$ associated to the primary constraints $\phi_k$, such that $\md\phi_k=\Omega(\cdot,X_k)$.
This condition requires that the corresponding gradients and HVF of all constraints be well defined so that we can add them to built the HVF associated to the total Hamiltonian. In case a) above, this means that the gradients have to be well defined, and since we do not have a boundary contribution to $\Omega$, then well defined means there should not be boundary contributions to the gradients. This is precisely the case of Regge-Teitelboim. In case b), we have bulk constraints and possible boundary constraints. In this latter case, both the gradient and HVF will be defined on the boundary only, and the process follows just as if there was a (boundary) theory living there. The bulk constraints might be more complicated since we are now allowing for boundary contributions to the gradient to survive which means that there might be non-zero components, at the boundary, of the bulk HVF. This consistency conditions shall impose some conditions on the bulk (as in the usual Dirac algorithm), but there might be new conditions at the boundary. These are to be thought of as
imposing (possible) new boundary conditions for the bulk fields, and {\it not} as new constraints on the Boundary DOF. This is an important distinction that shall become clear
when we consider an example in the sections to follow.

First class constraints will also either be ``bulk" or ``boundary" and they will generate gauge transformation since the corresponding HVF are well defined and tangent to the constraint surface. In this case, as we shall see in the examples to come, there might be contributions from boundary terms from the bulk part and from the boundary degrees of freedom, that might combine at the boundary. The corresponding HVF should be a degenerate direction of the total (pre-)symplectic structure that has both a bulk and a boundary contribution as in (\ref{omega-with-boundary}).

%Possible extension of the RTD algorithmBDOF
In what follows we illustrate the general theory developed in this section and perform a complete analysis of the 4-dimensional Maxwell + Pontryagin theory in a spacetime region with boundary, from different perspectives. We will show that in one of the approaches the usual RT criteria are sufficient for a consistent description of the theory, but in the other approach they are no longer valid. The resulting equations of motion in the bulk and boundary conditions are nevertheless the same, as expected.

\section{Hamiltonian analysis: Maxwell + Pontryagin}
\label{Sec:3}

This section has three parts. In the first one we introduce the 4-dimensional action principle and the corresponding 3+1 decomposition.
In the second part we perform the standard canonical analysis of the sistem. In the third one we use the geometrical approach and compare with the
standard Dirac analysis.
%\subsection{$3+1-$ decomposition}

Let us start from the action of the Maxwell theory coupled to a topological Pontryagin term, in a fixed background spacetime $(\mathcal{M}, g_{ab})$. We assume $\mathcal{M}$ to be a four dimensional smooth manifold with metric $g_{ab}$ of signature $(-,+,+,+)$.  Furthermore, we will take $\mathcal{M}$, or at least a portion of it $M\subseteq\mathcal{M}$, to be globally hyperbolic and such that it may be foliated as $M\approx
%\mathbb{R}
I\times\Sigma$, with $I=[t_1,t_2]$ and $\Sigma$ a spatial hypersurface. We will consider the case where $M$ has boundaries, namely is bounded by two Cauchy surfaces $\Sigma_1$ and $\Sigma_2$ and  a time-like boundary $\mathcal{D}=I\times\partial{\Sigma}$, so that $\partial{M}=\Sigma_1\cup\Sigma_2\cup\mathcal{D}$. The action is given by

\begin{equation}
S_{\rm MP} [A] = S_{\rm M} [A] + S_{\rm P} [A]= -\frac{1}{4} \int_{M} \md^{4} x\, \sqrt{ | g |} g^{ac} g^{bd} F_{cd} F_{ab}  - \frac{\theta}{4} \int_{M} 
\md^{4} x\,\tilde{\varepsilon}^{abcd} F_{ab} F_{cd} \, ,
\end{equation}
where $F_{ab}=\nabla_a A_b -\nabla_b A_a$ is the field strenght, $\tilde{\varepsilon}^{abcd}$ is the Levi-Civita symbol (Levi-Civita tensor density of weight 1) and $\theta$ is a coupling parameter.

The first step in the Hamiltonian analysis is to make a $3+1-$decomposition of the action, so we begin by decomposing the metric, $g^{ab} = h^{ab} - n^{a} n^{b}$, where $h_{ab}$ is the induced metric on the Cauchy slice, $\Sigma$, 
and $n^{a}$ is its unit normal. We introduce an everywhere timelike vector field $t^{a}$ and a ``time'' function $t$ such that hypersurfaces 
$t = {\rm const.}$ are diffeomorphic to $\Sigma$ and  $t^{a} \nabla_{a} t = 1$. 
%\footnote{This relation says that $t$ is the affine parameter of the integral curves along $t^{a}$} 
As usual, we can expand this vector field $t^{a}$ in terms of lapse $N$ and shift functions $N^a$,
\begin{equation}\label{Definition-t}
t^{a} = N n^{a} + N^{a}\, .
\end{equation}
%Also we have $\sqrt{|g|} = N \sqrt{ h}$. 

%\subsubsection{Maxwell}

%\begin{equation}\label{Maxwell}
%S_{M} [A] = \frac{\alpha}{2} \int_{\mathcal{M}} g^{a c} g^{bd} F_{cd} F_{ab} \sqrt{ | g |}\, \md^{4} x 
%\end{equation}
%From $g^{a b} = h^{a b} - n^{a} n^{b}$ and the definition (\ref{Definition-t}),
%\begin{equation}\label{definition-t}
%t^{a} =  N n^{a} + N^{a} \Rightarrow n^{a} = \frac{t^{a}- N^{a}}{N}\, ,
%\end{equation}
Now, since,
\begin{equation}
g^{a b} = h^{a b} - n^{a} n^{b} = h^{a b} - \frac{1}{N^{2}} (t^{a}- N^{a})(t^{b}- N^{b})\, ,  
%= h^{a b} - \frac{1}{N^{2}} [t^{a} t^{b} - t^{a} N^{b} - N^{a} t^{b} - N^{a} N^{b} ]
\end{equation}
it follows that,
%\begin{eqnarray}
%g^{a c} g^{b d} F_{ab} F_{cd} &=& h^{ac} h^{bd} F_{ab} F_{cd}  - 2 h^{ac} \frac{n^{b}}{N} (t^{d} F_{cd} - N^{d}F_{cd}) F_{ab}\nonumber \\ 
%&=& h^{ac} h^{b d} F_{a b} F_{cd} - \frac{2 h^{ac}}{N^{2}}( t^{d} F_{cd} - N^{d}F_{cd})( t^{b} F_{ab} - N^{b}F_{ab})\, .
%\end{eqnarray}
\be
g^{a c} g^{b d} F_{ab} F_{cd} 
%&=& h^{ac} h^{bd} F_{ab} F_{cd}  - 2 h^{ac} \frac{n^{b}}{N} (t^{d} F_{cd} - N^{d}F_{cd}) F_{ab}\nonumber \\ 
= \bigl[\,h^{ac} h^{b d}  - \frac{2 h^{ac}}{N^{2}}( t^{b} - N^{b})( t^{d}- N^{d})\, \bigr] F_{ab}F_{cd}\, .
\ee

Using Cartan's identity $\pounds_{t} A=t\cdot\md A+\md (t\cdot A)=t\cdot F + \md (t\cdot A)$, we obtain
\begin{equation}\label{LieDerivative}
%\pounds_{t} A=t\cdot\md A+\md (t\cdot A)=t\cdot F + \md (t\cdot A)
%\pounds_{t} A_{a} = t^{b} \nabla_{b} A_{a} + (\nabla_{a} t^{b}) A_{b}  = t^{b} \underbrace{\left[ \nabla_{b} A_{a} - 
%\nabla_{a} A_{b} \right]}_{F_{b a}} - \nabla_{a} ( t \cdot A ) 
t^{b} F_{ba} = \pounds_{t} A_{a} - \nabla_{a} \phi\, , 
\end{equation}
where $\phi:= t\cdot A$. 

Let us focus first on the Maxwell action. Taking (\ref{LieDerivative}) into account, as well as  $\sqrt{|g|} = N \sqrt{ h}$,
we can rewrite the Maxwell action as
\begin{eqnarray}
S_{\rm M}[A] &=& -\frac{1}{4} \int_{I} \md\, t \int_{\Sigma} \md^{3}x\, N \sqrt{h} \left\{ h^{ac} h^{bd} F_{ab} F_{cd} \right.\nonumber\\ 
&&  \left. - \frac{2 h^{ac}}{N^{2}} \left[ (\pounds_{t} A_{c} - \nabla_{c} \phi + N^{d} F_{cd})(\pounds_{t} A_{a} - 
\nabla_{a} \phi + N^{b} F_{ab}) \right]  \right\}\, . \label{MaxwellAction}
\end{eqnarray}
Note that, since both the one-form $\mathbf{A}$ 
and the field strength $\mathbf{F}$ are contracted with purely `spatial' objects ($n^{a} N_{a} = n^{a} h_{ab} = 0$), they are the pull-back to $\Sigma$ of the space-time objects. Nevertheless, we shall continue to use the same symbol, within the abstract-index notation to denote objects '
`living' in $M$, as well as their pull-backs to $\Sigma$. 

%The canonical momenta,
%\begin{equation}
%\Pi^{a}_{M} := \frac{\delta \mathcal{L}_{M}}{\delta (\pounds_{t} A_{a})} = - \frac{2\alpha \sqrt{h}}{N}\, h^{a c} (\pounds_{t} A_{c} - 
%\nabla_{c} ( t \cdot A ) + N^{d} F_{cd}).
%\end{equation}

%\subsubsection{Pontryagin}

On the other hand, since $\tilde{\varepsilon}^{abcd} = 4t^{[a} \tilde{\varepsilon}^{ bcd]} $ (see (\ref{A8}) in the Appendix A),
%(Romano, 1993), 
the Pontryagin term takes the form
\begin{equation}
S_{\rm P}[A] = -\theta \int_{\mathcal{M}} \md^{4} x\, \tilde{\varepsilon}^{abc} (t^d F_{da})\, F_{bc} 
=  -\theta \int_{\mathcal{M}}\md^4 x\,\tilde{\varepsilon}^{abc} F_{bc} \left( \pounds_{t} A_{a} - \nabla_{a}\phi\right)\, .
\label{PontryaginTerm}
\end{equation}
The canonical action of the Maxwell +  Pontryagin theory is the sum of (\ref{MaxwellAction}) and (\ref{PontryaginTerm}). In the following we will perform the canonical analysis of this theory.

\subsection{Dirac Analysis}
\label{Sec:3.A}

In this part we shall provide the standard Dirac analysis of the constrained system. We will show that there is no contribution to the momentum map from the boundaries and hence the standard Regge-Teitelboim criteria for the differentiability of the phase-space functions should be applied in order to have well defined canonical description of the theory.
%where the strategy on the boundaries is that as described in the previous section.

From (\ref{MaxwellAction}) and (\ref{PontryaginTerm}) we can calculate the corresponding canonical momenta of the full theory 
\begin{equation}\label{TotalMomenta-bulk}
{\tilde \Pi}^{a} := \frac{\delta L_{MP}}{\delta (\pounds_{t} A_{a})} =  \frac{ \sqrt{h}}{N}\, h^{a c} (\pounds_{t} A_{c} - 
\nabla_{c}\phi + N^{d} F_{cd}) - \theta\, \tilde{\varepsilon}^{abc} F_{bc}\, ,
\end{equation}
where $L_{MP}$ is the Lagrangian of the Maxwell-Pontryagin theory.
This implies that we can express the velocities in terms of momenta and fields as
\begin{equation}\label{LieDerivative-TotalMomentaBulk}
\pounds_{t} A_{a} = \frac{N}{ \sqrt{h}}\, h_{ab} \left(\tilde{\Pi}^{b} + \theta\,  \tilde{\varepsilon}^{bcd} F_{cd}  \right)
+ \nabla_{a}\phi - N^{b} F_{ab}\, .
\end{equation}
%This expression will be useful in the Legendre transform.

%\subsection{The Hessian and the primary constraints}

From the definition of the momenta (\ref{TotalMomenta-bulk}) we can see that there is only one primary constraint
\begin{equation}\label{primary_constraint}
\tilde{\Pi}_\phi := \hat{t}_{a} \tilde{\Pi}^{a} \approx 0\, ,
%\,\,\,\, {\rm or} \,\,\,\, 
%\tilde{\Pi}_\phi [\lambda] := \int \md^{3}x\, \lambda (x) n_{a} \tilde{\Pi}^{a} (x) \approx 0\, ,
\end{equation}
where $\hat{t}_a:=\nabla_a t=-\frac{1}{N}n_a$. 
%In order to determine the number of independent primary constraints we can calculate the nullity of the Hessian matrix,
%\begin{equation}
%H^{ab} = \frac{\delta^2 \mathcal{L}_{MP} }{\delta(\pounds_{t} A_{a} ) \delta(\pounds_{t} A_{b} ) }\, .
%\end{equation}
%Since the Pontryagin term is linear on $\pounds_{t} A_{a}$ it does 
%not contribute to the rank of the Hessian that is 3 (only due to the contribution of the Maxwell term). Therefore the nullity is 1 and The only 
%primary constraint is $\tilde{\Pi}_\phi [\lambda]$.

Now, the canonical action takes the form 
\begin{equation}\label{CanonicalActionMP}
S_{\rm MP}[A,\tilde{\Pi}]= \int_{I} \md\, t \int_{\Sigma} \md^{3}x\, \{ (\pounds_t \phi)\, \tilde{\Pi}_\phi + (\pounds_t A_a)\, \tilde{\Pi}^a- \mathcal{H}_{CMP} [A,\tilde{\Pi}]\} \, ,
\end{equation}
where the canonical Hamiltonian $H_{\rm CMP}=\int_{\Sigma}  \md^{3}x\, \mathcal{H}_{\rm CMP}$ is given by 
%\subsection{The canonical Hamiltonian and secondary constraints}
\begin{eqnarray}\label{CanonicalHamiltonian-TotalActionBulk}
\nonumber H_{\rm CMP}(A,\tilde{\Pi}) &=& \int_{\Sigma} \md^{3}x\, \left[ \frac{  \sqrt{h}}{4} N h^{ac} h^{bd} F_{ab} F_{cd} 
+ (\nabla_a\phi -N^b F_{ab})\tilde{\Pi}^a  \right.\\
%\nonumber 
&& \left. + \frac{N}{2 \sqrt{h}} h_{ab} (\tilde{\Pi}^{a}+\theta\,\tilde{\varepsilon}^{acd}F_{cd}) 
(\tilde{\Pi}^{b}+\theta\,\tilde{\varepsilon}^{bkn}F_{kn})\right]\, .
\end{eqnarray}
From the form of the canonical action we see that the momentum map has no contributions from the boundary $\partial\Sigma$.

The total Hamiltonian is  
\begin{equation}\label{TotalH-MP}
H_{\rm TMP} (A,\tilde{\Pi})=H_{\rm CMP} (A,\tilde{\Pi})+\int_{\Sigma} \md^{3}x\, u\,\tilde{\Pi}_{\phi}\, .
\end{equation}
and its variation is the sum of a bulk term and a boundary term
\begin{equation}
\delta  H_{\rm TMP} = V_{\mathrm{Bulk}}+ V_{\mathrm{Bound}}\, ,
\end{equation}
where
\[
V_{\mathrm{Bulk}} = \int_{\Sigma} \md^{3}x\,\bigl[ (\nabla_a \tilde{Q}^{ab}) \,\delta A_b + S_a\,\delta\tilde{\Pi}^{a}
- (\nabla_a\tilde{\Pi}^{a})\,\delta\phi + u\,\delta\tilde{\Pi}_{\phi}\bigr]\, ,
\]
and
\begin{equation}
V_{\mathrm{Bound}}=\int_{\partial\Sigma} \md^{2}y\,  r_a (\,\tilde{Q}^{ba}\, \delta A_b + \tilde{\Pi}^{a}\,\delta\phi\, )\, ,\label{VarBoundaryMP}
\end{equation}
where $r^a$ is the exterior unit normal to the boundary, such that $r^a n_a =0$ and $\tilde{\varepsilon}^{ab}=\tilde{\varepsilon}^{abc}r_c$. $\tilde{Q}^{ab}$ is an antisymmetric tensor density of weight 1, given by
\be
\tilde{Q}^{ab} := N^b\tilde{\Pi}^{a}-N^a\tilde{\Pi}^{b}-N\sqrt{h}F^{ab}-2\theta\frac{N}{\sqrt{h}}{\tilde{\epsilon}^{ab}}\,_c\, (\tilde{\Pi}^c+\theta\tilde{\epsilon}^{ckn}F_{kn})\, ,
\ee
and $S_a$ is a one-form given by
\be
S_a := \frac{N}{\sqrt{h}}(\tilde{\Pi}_{a}+\theta\tilde{\epsilon}_{acd}F^{cd})
+\nabla_a\phi -N^b F_{ab}\, .
\ee

We see that $H_{{\rm TMP}}$ is differentiable for arbitrary boundary variations $\delta A_b$ and $\delta\phi$ if
the following boundary conditions are satisfied 
\begin{eqnarray}
r_a  \tilde{Q}^{ab}       &=& 0\ \vert_{\partial\Sigma}\, ,\label{BCond1}\\
r_a\tilde{\Pi}^a &=& 0\ \vert_{\partial\Sigma}\, .\label{BCond2}
\end{eqnarray}
These are the most general boundary conditions that are necessary for the vanishing of the  boundary term $V_{\mathrm{Bound}}$, therefore providing the differentiability of the total Hamiltonian.  

Note that when some particular boundary conditions are imposed from the beginninig, $V_{\mathrm{Bound}}$ could vanish, due to the restriction on the boundary variations, even if (\ref{BCond1}) and (\ref{BCond2}) are not satisfied. This is the case with {\it Perfect conductor boundary conditions} (PCBC). As the name indicates, here we impose that the boundary behaves as a perfect conductor. In terms of the canonical variables this means,
\begin{eqnarray}
\phi &=& 0  \ \vert_{\partial\Sigma}\, ,\label{IIIset}\\
\varepsilon_{abc} r^b A^c &=& 0 \ \vert_{\partial\Sigma}\, . \label{IVset}
\end{eqnarray}
For these conditions $\delta\phi =0$, $\nabla_a\phi=0$ and $A^a\sim r^a$ on the boundary, so that the variations of the fields on the boundary are restricted: $\delta\phi =0$ and $\delta A_a\sim r_a$. Then, $V_{\mathrm{Bound}}$ vanishes.

In the following we will suppose that the boundary conditions (\ref{BCond1}) and (\ref{BCond2}) are satisfied, such that the expression (\ref{VarBoundaryMP}) vanishes. Then the total Hamiltonian is a differentiable functional and
we can continue with checking the consistency condition of the primary constraint, namely
\begin{equation}
\pounds_t{\tilde{\Pi}}_{\phi} := \{   \tilde{\Pi}_{\phi}, H_{{\rm TMP}}   \} \approx 0\, .
\end{equation}
%\rc{where $H_T$ is given by (\ref{TotalH-MP}).}
%\begin{equation}
%H_{T} := H_{C} + \int_{\Sigma} \md^{3}x\, u\, \tilde{P}_{\phi}\, .
%\end{equation}
This consistency condition leads to a new constraint, the Gauss constraint,
\begin{equation}\label{secondary_constraint}
\chi := \nabla_a \tilde{\Pi}^a\approx 0\, .
\end{equation}
The variation of the smeared Gauss constraint $\chi[w]=\int_{\Sigma} \md^{3}x\, w\,  (\nabla_a \tilde{\Pi}^a)$ is of the form
\begin{equation}
\delta\chi [w]= -\int_{\Sigma} \md^{3}x\,   (\nabla_a  w)\,\delta\tilde{\Pi}^a
+ \int_{\partial\Sigma} \md^{2}y\, w\, r_a\delta\tilde{\Pi}^a \, .
\end{equation}
It is differentiable if the boundary term vanishes, and that happens for all variations consistent with (\ref{BCond2}), since $r_a\delta\tilde{\Pi}^a = 0 \vert_{\partial\Sigma}$.
%This condition is fulfilled for arbitrary variations $\delta\Pi^a$ if $w\vert_{\partial\Sigma}=0$. 
Then,
%\begin{equation}
%\delta\chi [w]= -\int_{\Sigma} \md^{3}x\,   (\nabla_a  w)\,\delta\tilde{\Pi}^a\, .
%\end{equation}
the consistency condition  of the Gauss constraint  leads to
\begin{equation}
\{  \chi [w], H_{\rm TMP}\} = \int_{\Sigma} \md^{3}x\, \bigl[\, -\nabla_b (\, \tilde{Q}^{ab}\, \nabla_a\omega\, )
+ \tilde{Q}^{ab}\nabla_b\nabla_a\omega\, \bigr]=
\int_{\partial\Sigma} \md^{2}y\,  (r_b \tilde{Q}^{ba}) (\nabla_a\omega )=0\, ,
\end{equation}
since $\tilde{Q}^{ab}$ is antisymmetric and the boundary condition (\ref{BCond1}) is imposed. As a result, there are no tertiary constraints,
and the form and the algebra of the constraints are the same as in the Maxwell theory; they are not affected by the Pontryagin term. 

The corresponding equations of motion can be calculated through Poisson brackets as:
\begin{eqnarray}
 \pounds_t\phi &=& \{ \phi , H_{\rm TMP}\} = u\, ,\label{Heq1}\\
 \pounds_t\tilde{\Pi}_{\phi} &=& \{ \tilde{\Pi}_{\phi}, H_{\rm TMP}\} = \nabla_a{\tilde{\Pi}}^{a}\approx 0\, ,\label{Heq2}\\
 \pounds_t A_a &=& \{ A_a, H_{\rm TMP}\}= S_a \, ,\label{Heq3}\\
 %\{ A_a, H_{\rm TMP}\} =\frac{N}{\sqrt{h}}(\tilde{\Pi}_{a}+\theta\tilde{\epsilon}_{acd}F^{cd})\, 
%+\nabla_a\phi -N^b F_{ab} \, ,\label{Heq3}\\
 \pounds_t\tilde{\Pi}^{a} &=& \{ \tilde{\Pi}^{a}, H_{\rm TMP}\} = \nabla_b\tilde{Q}^{ab}\, .\label{Heq4}
 %-\nabla_a \bigl[2N^{[b}\tilde{\Pi}^{a]}-N\sqrt{h}F^{ab}-2\theta\frac{N}{\sqrt{h}}{\tilde{\epsilon}^{ab}}\,_c\, (\tilde{\Pi}^c+\theta\tilde{\epsilon}^{ckn}F_{kn})\bigr]\label{Heq4}
\end{eqnarray}

Let us now find the generator of gauge transformations, that is
 constructed as a linear combination of first class constraints
\begin{equation}\label{GaugeGeneratorMP}
G[\epsilon_1 ,\epsilon_2]=\int_{\Sigma}\md^3x\, ( \epsilon_1\tilde{\Pi}_\phi + 
\epsilon_2\nabla_a\tilde{\Pi}^a ) \, ,
\end{equation}
The generator is differentiable if $\epsilon_2\vert_{\partial\Sigma}=0$ or the condition (\ref{BCond2}) is fulfilled.
The corresponding gauge transformations are
\begin{eqnarray}
\delta\phi &=&  \epsilon_1\, , \ \ \ \ \  \delta\tilde{\Pi}_{\phi}=0\, ,\\ 
\delta A_a &=& -\nabla_a\epsilon_2 \, ,\ \ \ \delta\tilde{\Pi}^a=0\, . 
\end{eqnarray}
Full 4-dimensional $U(1)$ gauge symmetry is obtained for the choice  $\epsilon_1= \pounds_{t}\epsilon$ and $\epsilon_2=-\epsilon$, then
\begin{eqnarray}
 \delta_{\epsilon}\phi &=& \pounds_{t}\epsilon\, ,\nonumber\\
 \delta_{\epsilon} A_a &=& \nabla_a\epsilon\, .
\end{eqnarray}
Here we have used the result by \cite{Castellani:1981} that states that, in order to obtain (spacetime) gauge transformations,
one has to choose the parameter $\epsilon_1$ to be the time derivative of the parameter involved in spatial transformations.

In the previous analysis we have adopted the usual approach where the canonical theory is well defined only if all relevant structures are differentiable, in the sense that in their variation the boundary terms should vanish. As we have shown in the Section II this approach is completely justified since in this case the corresponding symplectic structure does not have a boundary contribution. In the following we will pursue the geometric point of view described in the Section II in order to demonstrate this equivalence.

\subsection{Geometric approach}
\label{Sec:3.B}

In this case the momentum map  is of the form
\begin{eqnarray}
\Pi_{\phi}[f] &=& \int_{\Sigma}\, \md^3x\, \tilde{\Pi}_{\phi}\,f\, ,\\
\Pi [V] &=& \int_{\Sigma}\, \md^3x\, \tilde{\Pi}^a\, V_a \, ,
%\int_{\partial\Sigma}\, \md^2x\, \tilde{P}^a_{\partial}\, v_a\, ,
\end{eqnarray}
%where in the surface integral $v_a$ is a pullback of the form $v$ to $\partial{\Sigma}$.
since $\pounds_t A$ appears only in the bulk term of the canonical action. 

In order to define the consistent evolution we need to find the Hamiltonian vector field $X_H$, tangent to the constrained hypersurface, such that
\begin{equation}\label{HamiltonianVFMP}
\ed H_{\rm TMP} (Y) = \Omega_{\rm MP} (Y, X_H) \, ,
\end{equation}
where $H_{\rm TMP}$ is given by (\ref{TotalH-MP}) and $\Omega_{\rm MP}$ is the symplectic structure on the kinematical phase space given by 
\be\label{SymplStructMP}
\Omega_{\rm MP}=\int_\Sigma\md^3x\,\, \bigl( \, \ed\tilde{\Pi}_\phi\wedge \ed\phi +  \ed\tilde{\Pi}^a\wedge \ed A_a\, \bigr) \, .
\ee
When acting  on the HVF whose  components are $X=(X^{\phi},X^{A}_a,X^{\Pi_\phi}, X^{\Pi\, a})$, we get
\begin{equation}
\Omega_{\rm MP} (X,Y) =
 \int_{\Sigma}\, \md^3x\, (X^{\Pi_\phi} Y^{\phi} - X^{\phi} Y^{\Pi_\phi}   + X^{\Pi\, a} Y^{A}_a - X^A_a Y^{\Pi\, a})
 \, .
\end{equation}
%where $X$, $Y$ are tangent to the phase space, with components $X=(X^{\phi},X^{A}_a,X^{\Pi_\phi}, X^{\Pi\, a})$.
Now the equation (\ref{HamiltonianVFMP}) takes the form
\begin{eqnarray}
& &\int_{\Sigma}\, \md^3x\, \bigl[ (-\nabla_a{\tilde{\Pi}}^{a})\, Y^{\phi}+ u\, Y^{\Pi_\phi}
+ S_a\, Y^{\Pi\, a}
+ (\nabla_a \tilde{Q}^{ab})\, Y^A_b\bigr]\notag\\
&  & +\int_{\partial\Sigma}\, \md^2y\, \bigl[ (r_a\tilde{Q}^{ba})\, Y^A_b + (r_a \tilde{\Pi}^{a})\,Y^\phi\,\bigr]\\
&= & \int_{\Sigma}\, \md^3x\, \bigl[ (X_H)^{\phi}\, Y^{\Pi_\phi}- (X_H)^{\Pi_\phi}\, Y^{\phi}   + {(X_H)}^A_{a}\, Y^{\Pi\, a}
-(X_H)^{\Pi\, a}\, Y^{A}_a \bigr]\, 
\notag
\label{HamVFMP1}
\end{eqnarray}
where in the integral over $\partial\Sigma$ we use the same notation for the pullback or restriction of the corresponding components to the boundary.
Now, comparing both sides of this equation we can read off the components of the Hamiltonian vector field $X_H$, in the bulk. From them, we obtain the Hamiltonian equations of motions in the bulk, the same as in Eqs.\, (\ref{Heq1}) - (\ref{Heq4}), namely 
\begin{eqnarray}
 \pounds_t\phi := {(X_H)}^{\phi} &=&  u\, ,\label{Heq5}\\
 \pounds_t \tilde{\Pi}_\phi := {(X_H)}^{\Pi_\phi} &=& \nabla_a{\tilde{\Pi}}^{a}\, ,\label{Heq6}\\
 \pounds_t A_a := {(X_H)}^A_a &=&  S_a\, ,\label{Heq7}\\
\pounds_t\tilde{\Pi}^a := {(X_H)}^{\Pi\, a}  &=&  \nabla_b \tilde{Q}^{ab}\, .\label{Heq8}
\end{eqnarray}
On the other hand, since on the RHS of the equation (\ref{HamVFMP1}) there are no boundary terms, all the boundary terms on the LHS must vanish. These are exactly the same boundary conditions on the bulk variables
obtained in the previous standard Dirac approach, given by (\ref{BCond1}) and  (\ref{BCond2}).
%\begin{equation}
% V_{{\rm Bound}} = 0,
% \end{equation}
%where $V_{{\rm Bound}}$ is given in (\ref{VarBoundaryMP}).

The consistency condition of the primary constraint, $\tilde{\Pi}_\phi\approx 0$, leads to the Gauss constraint, 
\begin{equation}
\pounds_t\tilde{\Pi}_\phi = {(X_H)}^{\Pi_\phi} = \nabla_a{\tilde{\Pi}}^{a} = 0\, .
\end{equation}
Its consistency condition is  satisfied since
\begin{equation}\label{TorsionFree}
 \pounds_t (\nabla_a{\tilde{\Pi}}^{a}) =\nabla_a {(X_H)}^{\Pi\, a}= \frac{1}{2}[\nabla_a ,\nabla_b]\,\tilde{Q}^{ab}  = 0\, .
\end{equation}
%due to torsion free condition. 
Hence, there are no tertiary constraints. 

To end this part, let us find the Hamiltonian vector field, $X_G$, of the generator of gauge transformations $G[\epsilon_1, \epsilon_2]$, given in (\ref{GaugeGeneratorMP}). We know that
\begin{equation}
\ed G (Y) =\Omega_{\rm MP} (Y, X_G)\, .
\end{equation}
Explicitly written it gives
\begin{eqnarray}
 & &\int_{\Sigma}\, \md^3x\, \bigl[\epsilon_1 Y^{\Pi_\phi} - (\nabla_a\epsilon_2)\, Y^{\Pi\, a} \bigr]
 +\int_{\partial\Sigma}\, \md^2y\, \epsilon_2\, r_a\, Y^{\Pi\, a}\notag\\
 &=&
 \int_{\Sigma}\, \md^3x\, \bigl[\, {(X_G)}^{\phi} Y^{\Pi_\phi}- {(X_G)}^{\Pi_\phi} Y^{\phi}   + {(X_G)}^A_a Y^{\Pi\, a}
-{(X_G)}^{\Pi\, a} Y^{A}_a \bigr]\, .
\end{eqnarray}
The boundary term on the LHS vanishes whenever the condition (\ref{BCond2})
%$r_a\tilde{\Pi}^a=0\vert_{\partial\Sigma}$, 
is satisfied, since from  $r_a\tilde{\Pi}^a=0\vert_{\partial\Sigma}$ it follows that $r_a\delta\tilde{\Pi}^a=r_a\, Y^{\Pi\, a}=0\vert_{\partial\Sigma}$.
In that  case there are no conditions on $\epsilon_2$ on the boundary. In the case of PCBC $\tilde{\Pi}^a$ is not restricted on the boundary,  and we need to impose an additional boundary condition $\epsilon_2=0\vert_{\partial\Sigma}$, so that the boundary term on the LHS vanishes. These are the same conditions that we have obtained in the previous approach. 
As a result the components of $X_G$ in the bulk are given by
\begin{eqnarray}
{(X_G)}^{\phi} &=&  \epsilon_1\, , \ \ \ \ \ \ \ \ \  {(X_G)}^{\Pi_\phi}=0\, ,\\ 
{(X_G)}^A_a &=& -\nabla_a\epsilon_2 \, ,\ \ \ {(X_G)}^{\Pi\, a}=0\, . 
\end{eqnarray}
Let us show that  this vector is a degenerate direction of the symplectic structure.
%, since for every vector $Z$ tangent to the constraint surface
%\begin{equation}
% \Omega (X_G, Z)=0\, .
%\end{equation}
On the constraints' surface we have $\delta\Pi_\phi =0$ and $\delta(\nabla_a\tilde\Pi^a)=0$. It follows that the components of every vector tangent to that surface, $Z$, have to satisfy
\begin{equation}\label{ConditionsZ}
 Z^{\Pi_\phi} =0\, , \ \ \ \ \nabla_a Z^{\Pi\, a}=0\, ,
 \end{equation}
 while the other components are arbitrary. Now, the pullback of $\Omega_{\rm MP}$ to the constraints' surface, $\Omega_{\rm MP\, g}$, is given by
\begin{eqnarray}
 \Omega_{\rm MP\, g} (X_G, Z) &=&\int_{\Sigma}\, \md^3x\, \bigl[\, {(X_G)}^{\Pi_\phi}\, Z_\Phi - {(X_G)}^{\phi}\, Z^{\Pi_\phi} + {(X_G)}^{\Pi\, a}\, Z^A_{a} -
 {(X_G)}^A_a\, Z^{\Pi\, a}\bigr] \\
 &=& \int_{\Sigma}\, \md^3x\, (\nabla_a\epsilon_2)\, Z^{\Pi\, a} =  -\int_{\Sigma}\, \md^3x\, \epsilon_2 (\nabla_a Z^{\Pi\, a}) +  \int_{\partial\Sigma} \md^2y\, \epsilon_2\, (r_a Z^{\Pi\, a}) = 0\, .\notag
\end{eqnarray}
We see that this expression vanishes due to (\ref{ConditionsZ}) and for the same boundary conditions that are necessary for the existence of $X_G$, that implies $r_aZ^{\Pi\, a}=0\vert_{\partial\Sigma}$ or  $\epsilon_2=0\vert_{\partial\Sigma}$ in the case of PCBC.

Let us end this part with a remark regarding boundary conditions. We are interested in providing a formalism, that includes boundary conditions needed for a consistent formulation. We have started with an action principle for fields that do not satisfy any a-priory boundary conditions, and we have found what conditions need to be satisfied for consistency. We are not interested in exploring different physical boundary conditions to see whether they satisfy the conditions we have found. The only exception we have made is for the so-called perfect conducting boundary conditions PCBC as an example of boundary conditions that {\it do} satisfy the more general conditions. The other point one should remark is that one obtains a completely consistent formalism without the need to introduce extra boundary terms for the action, nor the imposition of new boundary DOF, as sometimes is done in the literature.  

\section{Hamiltonian analysis: Maxwell + Chern-Simons}
\label{Sec:4}

The Pontryagin and Chern-Simons terms are equivalent, the former is given as a bulk term, and the later as the corresponding term on a boundary. The canonical analysis shows that the constraints and their algebra differ in these two theories, but the degrees of freedom, gauge symmetries and observables are the same \cite{CV-P}. Here we shall consider the Maxwell theory a with Chern-Simons term, and analyse how the addition of this boundary term modifies the canonical description of the theory. 
As before, $M$ is bounded by two Cauchy surfaces $\Sigma_1$ and $\Sigma_2$ and also has a time-like boundary $\mathcal{D}=I\times\partial{\Sigma}$, so that $\partial {M}=\Sigma_1\cup\Sigma_2\cup\mathcal{D}$. In the Appendix A we  show that the Pontryagin term in the bulk is equivalent to the Chern-Simons term on $\mathcal{D}$, since the variation of fields on the Cauchy surfaces vanish. 

The action is given by
\begin{equation}
S_{\rm MCS} [A] = S_{\rm M} [A] + S_{\rm CS} [A]= -\frac{1}{4} \int_{M} \md^{4} x\, \sqrt{ | g |} g^{ac} g^{bd} F_{cd} F_{ab}  + \frac{\theta}{2} \int_{\mathcal{D}} \md^{3} x\, \tilde{\varepsilon}^{abc} A_a F_{bc}\, .\label{MCSAction}
\end{equation}
%Let us choose the coordinates such that the boundary corresponds to $r=r_B$, and $r^a$ is the unit external normal to $\partial\Sigma$, in $\Sigma$. 
The $3+1$ decomposition of the Maxwell action is given in (\ref{MaxwellAction}), and, using (\ref{A9}), the Chern-Simons term can be rewritten, in its canonical form, as \cite{CV-P}
\begin{equation}\label{ChernSimonsAction}
S_{\rm CS} [A] = \theta\int_I \md t\int_{\partial{\Sigma}} \md^{2} y\, \tilde{\varepsilon}^{ab}  [(\pounds_t A_a) A_b + F_{ab}\,\phi]\, .
\end{equation}
In the previous expression $A_a$ and $F_{ab}$ are  pullbacks to the boundary of the quantities that `live' in $\Sigma$ and $\phi$ is evaluated on $\partial\Sigma$. From now on we will use the notation $A_a^\partial$ and $\phi^\partial$ for these boundary fields.

Since the Chern-Simons term contains time derivative of the connection, the momentum map  has a contribution from the boundary, as in (\ref{momentum-map-boundary}). In this case we have
%From (\ref{MaxwellAction}) and (\ref{ChernSimonsAction}) we can obtain the corresponding canonical momenta of the theory, as a distribution given by
%\begin{equation}\label{TotalMomenta-MCS}
%{\tilde \Pi}^{a} =   {\tilde P}^{a}
%- \theta\, \tilde{\varepsilon}^{abc} r_c A_b \delta (r-r_B)\, ,
%\end{equation}
%where, as before,
%\begin{equation}
%{\tilde P}^{a} =  \frac{ \sqrt{h}}{N}\, h^{a c} (\pounds_{t} A_{c} - \nabla_{c}\phi + N^{d} F_{cd}) 
%\end{equation}
%Again, we have a primary constraint
%\begin{equation}
%{\tilde \Pi}_{\phi} = 0\, . 
%\end{equation}
%\subsection{Geometric approach}
%In this case canonical momenta are of the form
\begin{eqnarray}
P_{\phi}[f] &=& \int_{\Sigma}\, \md^3x\, \tilde{P}_{\phi}\,f + \int_{\partial\Sigma}\, \md^2x\, \tilde{P}_{\phi^\partial}\, f^\partial \, ,\\
P[v] &=& \int_{\Sigma}\, \md^3x\, \tilde{P}^a\, v_a + 
\int_{\partial\Sigma}\, \md^2x\, \tilde{P}^a_{\partial}\, v_a^\partial\, ,
\end{eqnarray}
where in the surface integral $v_a^\partial$ is a pullback of the form $v$ to $\partial{\Sigma}$, and $f^\partial =f\vert_{\partial\Sigma}$.

From (\ref{MaxwellAction}), (\ref{MCSAction}) and (\ref{ChernSimonsAction}) it follows
\begin{eqnarray}
\tilde{P}_{\phi} &=& 0\, ,\\
\tilde{P}^{a} &=&  \frac{ \sqrt{h}}{N}\, h^{a c} (\pounds_{t} A_{c} - \nabla_{c}\phi + N^{d} F_{cd}) \, ,\\
\tilde{P}_{\phi^\partial} &=& 0\, ,\\
\tilde{P}^a_{\partial} &=&  \theta\, \tilde{\varepsilon}^{ab} A_b^\partial\, .
\end{eqnarray}
The $\Gamma_{\rm Bulk}$ part of the phase space is parametrized by the coordinates $(\phi, \tilde{P}_\phi ; A_a, \tilde{P}^a)$, and $\Gamma_{\rm Bound}$ is parametrized by $(\phi^\partial , \tilde{P}_{\phi^\partial} ; A_a^\partial , \tilde{P}^a_{\partial} )$.

The theory has primary constraints, one in $\Sigma$ 
\be
\tilde{C} := \tilde{P}_{\phi} = 0\, ,\label{MCS-BulkPrimaryConstraint}
\ee
and three on the boundary $\partial\Sigma$
\begin{eqnarray}
\tilde{C}_\partial &:=& \tilde{P}_{\phi^\partial}=0\, ,\label{MCS-BoundaryPrimaryConstraint1}\\
\tilde{C}^a_{\partial} &:=& \tilde{P}^a_{\partial} - \theta\, \tilde{\varepsilon}^{ab} A_b^\partial = 0 \, .
\label{MCS-BoundaryPrimaryConstraint}
\end{eqnarray}

The canonical Hamiltonian is given by
\begin{equation}
H_{\rm CMCS}= P[\pounds_t A]+P_{\phi}[\pounds_t\phi]-L_{\rm MCS}= H_{\rm CM}+ H_{\rm CCS}\, ,
\end{equation}
where $H_{\rm CM}$ and $H_{\rm CCS}$ are the canonical Hamiltonians of the Maxwell theory and the Chern-Simons theory %(defined on the boundary $\partial\Sigma$), 
respectively, with
\be\label{HamCM}
H_{\rm CM}= \int_{\Sigma} \md^{3}x \left[  \frac{  \sqrt{h}}{4} N h^{ac} h^{bd} F_{ab} F_{cd} 
+ (\nabla_a\phi -N^b F_{ab})\tilde{P}^a  + \frac{N}{2  \sqrt{h}} h_{ab}\tilde{P}^{a} \tilde{P}^{b} \right]\, ,
\ee
and
\begin{equation}
H_{\rm CCS}= -\theta\int_{\partial{\Sigma}} \md^{2} x\, \tilde{\varepsilon}^{ab} F_{ab}^\partial\,\phi^\partial\, .\label{HamCCS}
\end{equation}

The total Hamiltonian of the theory is obtained by adding to $H_{\rm CMCS}$ a linear combination of primary constraints, leading to
\be
H_{\rm TMCS} = H_{\rm CMCS} + \int_{\Sigma}\, \md^3x\, u\, \tilde{P}_{\phi} + \int_{\partial\Sigma}\, \md^2x\, \bigl[ \mu \tilde{P}_{\phi^\partial} + \mu_a (\tilde{P}^a_{\partial} - \theta\, \tilde{\varepsilon}^{ab} A_b^\partial )\bigr] \, .
\ee

The phase space is equipped with the symplectic structure, that now has a boundary term,
\be\label{SymplStructMCS}
\Omega_{\rm MCS}=\int_\Sigma\md^3x\,\, \bigl(\, \ed\tilde{P}_\phi\wedge \ed\phi + \ed\tilde{P}^a\wedge \ed A_a\, \bigr) +  \int_{\partial\Sigma}\md^2y\,\, \bigl(\, \ed\tilde{P}_{\phi^\partial}\wedge \ed\phi^\partial + \ed\tilde{P}^a_{\partial}\wedge \ed A_a^{\partial}\, \bigr)\, .
\ee
When acting on the HVF tangent to the phase space $\Gamma =\Gamma_{\rm Bulk}\times\Gamma_{\rm Bound}$, whose components are $X=(X^{\phi},X^A_a,X^{\tilde{P}_{\phi}}, X^{\tilde{P}\, a}; X^{\phi^\partial},X^{A^\partial}_a, X^{\tilde{P}_{\phi^\partial}}, X^{\tilde{P}^{\partial}a} )$, we obtain
\begin{eqnarray}
\Omega_{\rm MCS} (Y,X) &=& \int_{\Sigma}\, \md^3x\, \bigl( Y^{\tilde{P}_{\phi}} X^{\phi} - Y^{\phi} X^{\tilde{P}_{\phi}} +  Y^{\tilde{P}\, a} X^A_a - Y^A_a X^{\tilde{P}\, a} \bigr)\notag\\
&+&\int_{\partial\Sigma}\, \md^2y\, \bigl( Y^{\tilde{P}_{\phi^\partial}} X^{\phi^\partial} - Y^{\phi^\partial} X^{\tilde{P}_{\phi^\partial}} + Y^{\tilde{P}^{\partial}a} X^{A^\partial}_a - Y^{A^\partial}_a X^{\tilde{P}^{\partial}a}\bigr)\, .
\end{eqnarray}
%where the vectors $X$ and $Y$ are tangent to the phase space $\Gamma =\Gamma_{\rm Bulk}\times\Gamma_{\rm Bound}$, with components $X=(X^{\phi},X^A_a,X^{\tilde{P}_{\phi}}, X^{\tilde{P}\, a}; X^{\phi^\partial},X^{A^\partial}_a, X^{\tilde{P}_{\phi^\partial}}, X^{\tilde{P}^{\partial}a} )$.
In order to define the consistent evolution we need to find the Hamiltonian vector field $X_H$, tangent to the constrained hypersurface, such that
\begin{equation}\label{HamiltonianVF}
\ed H_{\rm TMCS} (Y) = \Omega_{\rm MCS} (Y,X_H) \, ,
\end{equation}
that explicitly is of the form
\begin{eqnarray}\label{LongEq}
& &\int_{\Sigma}\, \md^3x\, \bigl[\, {(X_H)}^{\phi}Y^{\tilde{P}_{\phi}}  - {(X_H)}^{\tilde{P}_{\phi}} Y^{\phi}  +  {(X_H)}^A_a Y^{\tilde{P}\, a}  -  {(X_H)}^{\tilde{P}\, a} Y^A_a \, \bigr]\notag\\
&+&\int_{\partial\Sigma}\, \md^2y\, \bigl[ \, {(X_H)}^{\phi^\partial} Y^{\tilde{P}_{\phi^\partial}}  -{(X_H)}^{\tilde{P}_{\phi^\partial}} Y^{\phi^\partial} + {(X_H)}^{A^\partial}_a Y^{\tilde{P}^{\partial}a}  - {(X_H)}^{\tilde{P}^{\partial}a} Y^{A^\partial}_a \, \bigr]\notag\\
&=& \int_{\Sigma}\, \md^3x\, \bigl[\, u Y^{\tilde{P}_{\phi}}- (\nabla_a{\tilde{P}}^{a}) Y^{\phi} + (\nabla_b \tilde{B}^{ba})\, Y^A_a +
J_a\, Y^{\tilde{P}\, a}\,\bigr]\notag\\
&+& \int_{\partial\Sigma}\, \md^2y\, \bigl[\, r_b\, \tilde{B}^{ab}  + \theta\tilde{\varepsilon}^{ba}(2\nabla_b\phi^\partial -\mu_b)\,\bigr] Y^{A^\partial}_a 
+ \mu Y^{\tilde{P}_{\phi^\partial}} + {\mu_a}Y^{\tilde{P}^{\partial}\, a}  + (r_a \tilde{P}^{a}-\theta\tilde{\varepsilon}^{ab}F^\partial_{ab})\, Y^{\phi^\partial}\, \bigr]\, , \notag
\end{eqnarray}
where we have introduced the notation
\be
\tilde{B}^{ab} := -\sqrt{h}NF^{ab}+N^b{\tilde{P}}^{a}-N^a{\tilde{P}}^{b}\, ,
\ee
and
\be
J_a := \frac{N}{\sqrt{h}} {\tilde{P}}_{a}+ \nabla_a\phi -N^b F_{ab}\, .
\ee
In (\ref{HamiltonianVF}) we have made integration by parts in $\ \ed H_{\rm TM}(Y)$, where $H_{\rm TM}$ is the total Hamiltonian of the Maxwell theory. Since $A_a^\partial$ is a pullback of $A_a$ to $\partial\Sigma$ and $\phi^\partial = \phi\vert_{\partial\Sigma}$, we have also identified the corresponding components of the vector field $Y$ on the boundary.

Now  we can obtain the Hamiltonian equations of motion in  the bulk and on the boundary, as well as the boundary conditions for the bulk variables. 
Since $Y$ is arbitrary, let us first analyze the case when $Y\in T\Gamma_{\rm Bulk}$, such that $Y\vert_{\partial\Sigma}=0$. Then, from (\ref{LongEq}) we obtain that components of the Hamiltonian vector field in the bulk are
\begin{eqnarray}
\pounds_t\phi &:=& {(X_H)}^{\phi} = u\, ,\label{Xphi}\\
\pounds_t{\tilde{P}_\phi} &:=& {(X_H)}^{\tilde{P}_{\phi}} = \nabla_a{\tilde{P}}^{a}\, ,\label{Xpphi}\\
\pounds_t{A_a} &:=& {(X_H)}^A_a = J_a\, ,\label{Xa2}\\
\pounds_t\tilde{P}^a &:=& {(X_H)}^{\tilde{P}\, a} =  \nabla_b \tilde{B}^{ab}\, .\label{Xp}
\end{eqnarray}
On the other hand, for $Y\in T\Gamma_{\rm Bound}$, from (\ref{LongEq}) we obtain the equations of motion for boundary degrees of freedom
\begin{eqnarray}
\pounds_t\phi^\partial &:=& {(X_H)}^{\phi^\partial} = \mu\, ,\\
\pounds_t\tilde{P}_{\phi^\partial} &:=& {(X_H)}^{\tilde{P}_{\phi^\partial}} =-r_a \tilde{P}^a+\theta\tilde{\varepsilon}^{ab}F^\partial_{ab}\, ,\\
\pounds_t A_a^\partial &:=& {(X_H)}^{A^\partial}_a =  \mu_a\, ,\label{Xb2}\\
\pounds_t \tilde{P}^a_\partial &:=& {(X_H)}^{\tilde{P}^{\partial}a} = -r_b\tilde{B}^{ab}+\theta\tilde{\varepsilon}^{ab}(2\nabla_b\phi^\partial - \mu_b)\, .\label{Xa-boundary}
\end{eqnarray}

The consistency conditions of the bulk primary constraint, leads to the Gauss constraint in the bulk
\be\label{Gauss-bulk}
\pounds_t\tilde{C} =  {(X_H)}^{\tilde{P}_{\phi}} = \nabla_a{\tilde{P}}^{a}=0\, .
\ee
The consistency conditions of the boundary constraints are 
\begin{eqnarray}
\pounds_t\tilde{C}_{\partial} &=& {(X_H)}^{\tilde{P}_{\phi^\partial}} =-r_a \tilde{P}^a+ \theta\tilde{\varepsilon}^{ab}F_{ab} = 0\, \vert_{\partial\Sigma}\, ,
\label{BoundCond1}\\
\pounds_t\tilde{C}^a_{\partial} &=& {(X_H)}^{\tilde{P}^{\partial}a} + \theta\tilde{\varepsilon}^{ab}{(X_H)}^{A^\partial}_b  = 
 -r_b\tilde{B}^{ab}+2\theta\tilde{\varepsilon}^{ab}(\nabla_b\phi - \mu_b) =0\, \vert_{\partial\Sigma}\, .\label{mu}
\end{eqnarray}
The first consistency condition (\ref{BoundCond1}) is a boundary condition for bulk variables
\begin{equation}\label{BCond4}
 r_a \tilde{P}^a - \theta\tilde{\varepsilon}^{ab}F_{ab} = 0\, \vert_{\partial\Sigma}\, ,
\end{equation}
equivalent to the condition (\ref{BCond2}) previously obtained in Maxwell + Pontryagin theory. An important point to remark here is that the consistency
condition above is to be taken as a {\it boundary condition} and not as a boundary constraint since it involves bulk variables. If it were to include only
boundary variables then it would have to be regarded as a new boundary constraint. Not properly identifying this distinction can (and has) lead to confusion.

From the eq.$\,$(\ref{mu}) we can obtain the value for $\mu_a$,
\be \label{value_mu}
 \mu_a = -\frac{1}{4\theta}\, \underaccent{\tilde}{\varepsilon}_{ab}\, r_c \tilde{B}^{cb} + \nabla_a\phi\, \vert_{\partial\Sigma}\, ,
\ee
as a consequence of the second class nature of the boundary primary constraints (\ref{MCS-BoundaryPrimaryConstraint}). 

The continuity of ${(X_H)}^A_a$ implies that its pullback to the boundary is ${(X_H)}^{A^\partial}_a$, so we pullback (\ref{Xa2}) to the boundary and identify it with (\ref{Xb2}), and taking into account the value of $\mu_a$ given in (\ref{value_mu}) we obtain an additional boundary condition
\begin{equation}\label{BCond3}
r_a \tilde{B}^{ba} + 2\theta\, \tilde{\varepsilon}^{ba}\bigl(\frac{N}{\sqrt{h}}\tilde P_a - N^d F_{ad}\bigr)=0\, \vert_{\partial\Sigma}\, .
\end{equation}
In the next section we shall show that this expression  is equivalent to the boundary conditions given in (\ref{BCond1}).

Using the same argument as in (\ref{TorsionFree}) we obtain 
\begin{equation}
\pounds_t (\nabla_a{\tilde{P}}^{a}) =  \nabla_a {(X_H)}^{\tilde{P}\, a}= \frac{1}{2}[\nabla_a,\nabla_b ]\tilde{B}^{ab}= 0\, ,
\end{equation}
so there are no tertiary constraints. Now we have the full set of the constraints, the bulk ones are first class, while the boundary ones contain one first class and two second class constraints.

Let us now construct the generator of gauge transformations of Maxwell + Chern-Simons theory. As before, the generator is a linear combination of first class constraints, and in this case it also has a contribution from a boundary FCC, 
\begin{equation}\label{GaugeGeneratorMCS}
G[\epsilon_1 ,\epsilon_2, \epsilon_3]=\int_{\Sigma}\md^3x\, ( \epsilon_1\tilde{P}_\phi + 
\epsilon_2\nabla_a\tilde{P}^a ) + \int_{\partial\Sigma}\, \md^2y\,  \epsilon_3\tilde{P}_{\phi^\partial} \, .
\end{equation}
The corresponding Hamiltonian vector field, $X_G$, is obtained from
\begin{equation}
\ed G (Y) =\Omega_{\rm MCS} (Y, X_G)\, .
\end{equation}
Explicitly written it gives
\begin{eqnarray}\label{HVF-GMCS}
 & &\int_{\Sigma}\, \md^3x\, \bigl[\epsilon_1 Y^{\tilde{P}_\phi} - (\nabla_a\epsilon_2)\, Y^{\tilde{P}\, a} \bigr]
 +\int_{\partial\Sigma}\, \md^2y\, \bigl(\, \epsilon_2\,  r_a\, Y^{\tilde{P}\, a} + \epsilon_3\, Y^{\tilde{P}_{\phi^\partial}}\,\bigr) \notag\\
 &=&\int_{\Sigma}\, \md^3x\, \bigl[\, {(X_G)}^{\phi}Y^{\tilde{P}_{\phi}}  - {(X_G)}^{\tilde{P}_{\phi}} Y^{\phi}  +  {(X_G)}^A_a Y^{\tilde{P}\, a}  -  {(X_G)}^{\tilde{P}\, a} Y^A_a \, \bigr]\\
&+&\int_{\partial\Sigma}\, \md^2y\, \bigl[ \, {(X_G)}^{\phi^\partial} Y^{\tilde{P}_{\phi^\partial}}  -{(X_G)}^{\tilde{P}_{\phi^\partial}} Y^{\phi^\partial} + {(X_G)}^{A^\partial}_a Y^{\tilde{P}^{\partial}a}  - {(X_G)}^{\tilde{P}^{\partial}a} Y^{A^\partial}_a \, \bigr]\, .\notag
\end{eqnarray}
The boundary conditions impose conditions on the components of HVF. From (\ref{BCond4}) it follows that
\be\label{BoundY}
r_a\, Y^{\tilde{P}\, a} - 2\theta \tilde{\varepsilon}^{ab}\nabla_a Y^{A^\partial}_b=0\,\vert_{\partial\Sigma}\, .
\ee
Also, on the hypersurface of boundary primary constraints we have that for any vector $Y$ (including $X_G$), its boundary components satisfy 
\be\label{YBound}
Y^{\tilde{P}^{\partial}a}=\theta \tilde{\varepsilon}^{ab}Y^{A^\partial}_b\, ,
\ee
so that the boundary terms in (\ref{HVF-GMCS}) can be rewritten as
\begin{eqnarray}\label{HVF-GMCS_Bound}
 & &  \int_{\partial\Sigma}\, \md^2y\, \bigl[ -2\theta\tilde{\varepsilon}^{ab} (\nabla_a\epsilon_2)\, Y^{A^\partial}_b  + \epsilon_3\, Y^{\tilde{P}_{\phi^\partial}}\,\bigr] \notag\\
 &=&\int_{\partial\Sigma}\, \md^2y\, \bigl[ \, {(X_G)}^{\phi^\partial} Y^{\tilde{P}_{\phi^\partial}}  -{(X_G)}^{\tilde{P}_{\phi^\partial}} Y^{\phi^\partial} + 2\theta\tilde{\varepsilon}^{ab} {(X_G)}^{A^\partial}_a  Y^{A^\partial}_b \, \bigr]\, .
\end{eqnarray}
From (\ref{HVF-GMCS}) we can read off the components of $X_G$ in the bulk 
\begin{eqnarray}
{(X_G)}^{\phi} &=&  \epsilon_1\, , \ \ \ \ \ \ \ \ \  {(X_G)}^{\tilde{P}_\phi}=0\, ,\\ 
{(X_G)}^A_a &=& -\nabla_a\epsilon_2 \, ,\ \ \ {(X_G)}^{\tilde{P}\, a}=0\, . 
\end{eqnarray}
From (\ref{HVF-GMCS_Bound}) we obtain the components of $X_G$ on the boundary
\begin{eqnarray}
{(X_G)}^{\phi^\partial} &=&  \epsilon_3\, , \ \ \ \ \ \ \ \ \  {(X_G)}^{\tilde{P}_{\phi^\partial}}=0\, ,\\ 
{(X_G)}^{A^\partial}_a &=& -\nabla_a\epsilon_2 \, ,\ \ \ {(X_G)}^{\tilde{P}^\partial\, a}=-\theta \tilde{\varepsilon}^{ab}\nabla_b\epsilon_2\, ,
\end{eqnarray}
where the last component follows from (\ref{YBound}). We also have that $\epsilon_1=\epsilon_3\,\vert_{\partial\Sigma}$, since the boundary gauge transformation of $\phi$ and $A_a$  are of the same form as the corresponding bulk transformations. Again, for $\epsilon_1 =\pounds_t\epsilon$ and $\epsilon_2 =\epsilon$, we obtain the $U(1)$ gauge transformations.

Now, it is easy to see that $X_G$ is a degenerate direction of the symplectic structure $\Omega_{\rm MCSg}$, following the same arguments as in the Maxwell + Pontryagin case. %\rc{I have no time to check this, but it should be true.}
%, since for every vector $Z$ tangent to the constraint surface
%\begin{equation}
% \Omega (X_G, Z)=0\, .
%\end{equation}
%On the constraints' surface we have $\delta\Pi_\phi =0$ and $\delta(\nabla_a\tilde\Pi^a)=0$. It follows that 
The components of every vector tangent to the constraints' surface, $Z$, have to satisfy
\begin{equation}
 Z^{\tilde{P}_\phi} =0\, , \ \ \ \ \nabla_a Z^{\tilde{P}\, a}=0\, ,\ \ \ \ Z^{\tilde{P}^\partial_\phi} =0\, ,
 \end{equation}
 while the other components are arbitrary. Now, the pullback of $\Omega_{\rm MCS}$ to the constraints' surface, $\Omega_{\rm MCS\, g}$, is given by
%\begin{eqnarray}
% \Omega_{\rm MP\, g} (X_G, Z) &=&\int_{\Sigma}\, \md^3x\, \bigl[\, {(X_G)}^{\Pi_\phi}\, Z_\Phi - {(X_G)}^{\phi}\, Z^{\Pi_\phi} + {(X_G)}^{\Pi\, a}\, Z^A_{a} -
 %{(X_G)}^A_a\, Z^{\Pi\, a}\bigr] \\
 %&=& \int_{\Sigma}\, \md^3x\, (\nabla_a\epsilon_2)\, Z^{\Pi\, a} =  -\int_{\Sigma}\, \md^3x\, \epsilon_2 (\nabla_a Z^{\Pi\, a}) +  %\int_{\partial\Sigma} \md^2y\, \epsilon_2\, (r_a Z^{\Pi\, a}) = 0\, .\notag
%\end{eqnarray}
\begin{eqnarray}
 \Omega_{\rm MCS\, g} (X_G, Z) &=&
 %\int_{\Sigma}\, \md^3x\, \bigl[\, {(X_G)}^{\Pi_\phi}\, Z_\Phi - {(X_G)}^{\phi}\, Z^{\Pi_\phi} + {(X_G)}^{\Pi\, a}\, Z^A_{a} -\int_{\partial\Sigma} \md^2y\, 2\theta\tilde{\varepsilon}^{ab} (\nabla_a\epsilon_2)\,  Z^{A^\partial}_b 
 %{(X_G)}^A_a\, Z^{\Pi\, a}\bigr] \\
  \int_{\Sigma}\, \md^3x\, (\nabla_a\epsilon_2)\, Z^{\tilde{P}\, a} +2\theta\int_{\partial\Sigma} \md^2y\, \tilde{\varepsilon}^{ab} (\nabla_a\epsilon_2)\,  Z^{A^\partial}_b\\  
 &=&  -\int_{\Sigma}\, \md^3x\, \epsilon_2 (\nabla_a Z^{\tilde{P}\, a}) +  \int_{\partial\Sigma} \md^2y\, \epsilon_2\, \bigl(\, r_a Z^{\tilde{P}\, a}- 2\theta\tilde{\varepsilon}^{ab} \nabla_a Z^{A^\partial}_b \, \bigr) = 0\, ,\notag
\end{eqnarray}
and vanishes for an arbitrary vector $Z$, tangent to the constraints' surface that satisfies the boundary condition (\ref{BoundY}).

As we have shown, both descriptions of the physical system, as Maxwell + Pontryagin in Sec.~\ref{Sec:3} or as Maxwell + Chern-Simons in Sec.~\ref{Sec:4} are completely equivalent, as seen by the same equations of motion, gauge symmetries {\it and} boundary conditions. But this equivalence only holds if we adopt the formalism we described in Sec.~\ref{Sec:2} for field theories in regions with boundaries. That is, we need to allow for a contribution from the boundary to the symplectic structure 
{\it and} the gradient of functions (and thus transcending from the RT criteria), for the formalism to be fully consistent. In other words, the physical system under consideration can be seen as validating the formalism of Sec.~\ref{Sec:2}. Let us now see how we can connect these two description via a canonical transformation.

\section{Comparison}
\label{Sec:5}

%\subsection{Canonical transformation}
%\label{Sec:5.A}

In this section we shall show that the Maxwell + Pontryagian and Maxwell + Chern-Simons theories are related by a canonical transformation. In the case of a region
without boundaries, it is well known that one can go from the Maxwell to the Maxwell + Pontryagin theory through a canonical transformation \cite{Jackiw}. In the case of 
a region with boundary, the situation becomes more interesting, as boundary contributions ``appear" during the process. Let us start by recalling what it means to 
have a canonical transformation. As is the case for any diffeomorphism, one can adopt a {\it passive} or an {\it active} point of view. In the active viewpoint, one has a 
mapping acting on the phase space to itself, such that the {\it Lie dragged} two form $\bar{\Omega}$ coincides with the original two form $\Omega$. Diffeomorphisms
with that property are called symplectomorphisms and are generated, locally, by HVF. The passive viewpoint is to take the phase space fixed, together with 
$\Omega$, and to a consider change of canonical variables, from $(Q,P)$ to $(\bar{Q},\bar{P})$, such that the given symplectic form $\Omega$ takes the ``same" form
$\Omega=\md P\wedge\md Q = \md \bar{P}\wedge\md \bar{Q}$. In the case of field theory with boundaries, the situation becomes more subtle, as we shall now see. An important common theme in both viewpoints is that canonical transformations are generated by a {\it generating function}. Thus, if we are able to produce a well defined function that is responsible for generating the transformation on phase space, we can be sure that this transformation is indeed canonical. 

For our system under consideration, that implies that they have the same symplectic structure $\Omega$. As we shall see in what follows,  when $\Omega$ is written in terms of the Maxwell + Pontryagin phase space variables, it does not have any boundary terms, while they appear when we write down the expression for $\Omega$ in Maxwell + Chern-Simons phase space variables. That is, a well defined canonical transformation allows for a boundary term, apart from the canonical {\it Darboux} form in the bulk. Let us now see how that comes about.

It is well known that, starting from a given symplectic potential $\Theta$, one can add a gradient of some functional $\mathcal{G}$, without changing the symplectic structure of the theory,
\begin{equation}
\bar{\Theta} =\Theta +\ed \mathcal{G}\ \ \ \Longrightarrow\ \ \ \bar{\Omega} = \Omega \, . 
\end{equation}
The function $\mathcal{G}$ is precisely the generator of the canonical transformation. 
Let us now
start from the symplectic potential of the Maxwell + Pontryagin theory
\be
\Theta_{\rm MP}=\int_\Sigma\md^3x\, (\tilde{\Pi}_{\phi}\; \ed\phi + \tilde{\Pi}^a\; \ed A_a ) \, ,
\ee
and for $\mathcal{G}$ given by
\be
\mathcal{G} = \theta\int_\Sigma\md^3x\,\, \tilde{\epsilon}^{abc}A_a (\nabla_b A_c)\, ,
\ee
we obtain
\be
\bar{\Theta}_{\rm MP}  =\Theta_{\rm MP}+\ed \mathcal{G} = \int_\Sigma\md^3x\, (\tilde{P}_{\phi}\; \ed\phi +\tilde{P}^a\; \ed A_a) + \theta \int_{\partial\Sigma}\,\md^2y\, 
\tilde{\epsilon}^{ab}A_b\; \ed A_a\, ,
\ee
which has the standard form, but for new set of momenta, where
\begin{eqnarray}
\tilde{P}_{\phi} &=&  \tilde{\Pi}_{\phi}\, , \label{CT1}\\
\tilde{P}^a &=& \tilde{\Pi}^a + \theta\, \tilde{\epsilon}^{abc}F_{bc}\label{CT2}\, .
\end{eqnarray}
These are exactly the canonical momenta of the Maxwell theory of Sec.\ref{Sec:4}. 
The ``new" symplectic structure $\bar{\Omega}_{\rm MP}={\Omega}_{\rm MP}$ takes the form
\be
%\int_\Sigma\md^3x\,\, \ed\tilde{\Pi}^a\wedge \ed A_a =
\bar{\Omega}_{\rm MP}=
\int_\Sigma\md^3x\,\, \bigl(\,\ed\tilde{P}_{\phi}\wedge\ed\phi + \ed\tilde{P}^a\wedge \ed A_a\,\bigr) +\theta\, \int_{\partial\Sigma}\md^2y\, \tilde{\epsilon}^{ab}\, \ed A_b\wedge \ed A_a\, .
\label{symp-st-maxwell-plus-boundary}
\ee
As described above, what we are seeing is that the fixed symplectic two-form $\Omega$ has acquired a boundary contribution when written in terms of the
Maxwell ``coordinates", which means that the Maxwell + Pontryagin theory is {\it not} equal to the Maxwell theory since there are these extra terms at the boundary.
In principle we do not know what this boundary theory is, so one can try to find what it is. 
The first clue is that the boundary contribution to (\ref{symp-st-maxwell-plus-boundary}) takes precisely 
the form of the pullback of the symplectic structure of the Maxwell + Chern-Simons theory, $\Omega_{\rm MCS}$ (\ref{SymplStructMCS}), to the surface of second class boundary constraints defined by 
Eq.~(\ref{MCS-BoundaryPrimaryConstraint}). This hints to a Chern-Simons theory on the boundary. 
One might guess that one should also acquire a boundary term to, say, the Hamiltonian so that it coincides 
with the Chern-Simons theory. As we shall show the situation is more subtle than anticipated.
We will see that there is also a boundary contribution, but to the Hamiltonian vector fields, 
that confirms the suspicion that we have a description equivalent to the Maxwell +  Chern-Simons of Sec.~\ref{Sec:4}.

Let us start by considering the canonical Hamiltonian of the Maxwell + Pontryagin theory given by Eq.~(\ref{CanonicalHamiltonian-TotalActionBulk}). It is only after the canonical transformation that $\bar{H}_{\rm CMP}(\phi ,\tilde{P}_{\phi}; A_a,\tilde{P}^a)$ 
%\begin{eqnarray}\label{CanonicalHamiltonian-TotalActionBulk-CanonicalTransformation}
%\nonumber H_{C}(A,\tilde{P}) &=& \int_{\Sigma} \md^{3}x \left[  \frac{  \sqrt{h}}{4} N h^{ac} h^{bd} F_{ab} F_{cd} 
%+ (\nabla_a\phi -N^b F_{ab})\tilde{P}^a  \right.\\
%\nonumber 
%&& \left. + \frac{N}{2  \sqrt{h}} h_{ab}\tilde{P}^{a} \tilde{P}^{b} 
%-\theta\,\tilde{\varepsilon}^{acd}(\nabla_a\phi) F_{cd}\right]\, ,
%\end{eqnarray}
%and it can be \bc{alternatively} written as
%\begin{eqnarray}\label{CanonicalHamiltonian-TotalActionBulk-CanonicalTransformation-v2}
%\nonumber H_{C}(A,\tilde{P}) &=& \int_{\Sigma} \md^{3}x \left\{ N\left( \frac{\sqrt{h}}{4}  h^{ac} h^{bd} %F_{ab} F_{cd} 
%+ \frac{1}{2 \sqrt{h}} h_{ab}\tilde{P}^{a} \tilde{P}^{b} \right)-N^b F_{ab}\tilde{P}^a  \right.\\
%\nonumber 
%&& \left. -\phi\nabla_a \tilde{P}^a + \nabla_a\left[ \phi (\tilde{P}^a-\theta\,\tilde{\varepsilon}^{acd}%F_{cd})\right]\right\} \, ,
%\end{eqnarray}
%since $\tilde{\varepsilon}^{acd}\nabla_a F_{cd}=0$ (Bianchi identity).
becomes a sum of two terms, one of them is the canonical Hamiltonian of the Maxwell theory, $H_{\rm CM}$ (\ref{HamCM}), and the other term
is an additional contribution from the Pontryagin term $H_{\rm CP}$,
\begin{equation}
\bar{H}_{\rm CMP}  = H_{\rm CM}   + H_{\rm CP}\, ,
\end{equation}
where 
\begin{equation}\label{H_Pontryagin}
H_{\rm CP}(\phi ,A_a)=-\theta\int_{\Sigma} \md^{3}x\, \tilde{\varepsilon}^{acd}(\nabla_a\phi)\, F_{cd}   \, .
\end{equation}

The total Hamiltonian is  given by
\begin{equation}\label{TotalH-MP-Canonical}
\bar{H}_{\rm TMP} =\bar{H}_{\rm CMP} +\int_{\Sigma} \md^{3}x\, u\,\tilde{P}_{\phi}\, .
\end{equation}
In order to compare the results obtained in the Maxwell + Chern-Simons theory, with the corresponding ones we have obtained here in Maxwell + Pontryagin after the canonical transformation, we shall first construct  the evolutionary Hamiltonian vector field $\bar{X}_H$, as a solution of
\be
\ed \bar{H}_{\rm TMP} (Y)=\bar{\Omega}_{\rm MP} (Y, \bar{X}_H)\, .\label{ecuacionimportantecan-trans}
\ee
The integrals over $\Sigma$ in this expression give the bulk components of  $\bar{X}_H$, that are the 
same as (\ref{Xphi})-(\ref{Xp}). The new and subtle feature here is that, since we have a boundary 
contribution to the symplectic structure, then we have to adopt the strategy outlined in Sec.~\ref{Sec:2} 
for the case when there are boundary contributions. This means that, in constrast to the original Maxwell + Pontryagin case, we
will also have a corresponding boundary contribution from the 
gradients, and the HVFs. Thus, the boundary contribution to the previous equation (\ref{ecuacionimportantecan-trans}) is of the form
\begin{equation}
\int_{\partial\Sigma} \md^{2}y\, \bigl[ ( r_a\tilde{B}^{ba} -2
\theta \tilde{\varepsilon}^{ba}\nabla_ a\phi )\, Y^A_b + (r_a \tilde{P}^{a} -  \theta \tilde{\varepsilon}^{ab}F_{ab})\, Y^\phi\bigr] = 2\theta\int_{\partial\Sigma} \md^{2}y\, \tilde{\varepsilon}^{ab}{(\bar{X}_H)}^A_a\, Y^A_b
\, .\label{Vboundary}
\end{equation}
%where $r^a$ is the exterior unit normal to the boundary and $\tilde{\varepsilon}^{ab}=\tilde{\varepsilon}^{abc}r_c$. 
Thus, comparing both sides of this expression we again obtain the boundary condition (\ref{BCond4}), and the following components of $\bar{X}_H$ on the boundary
\begin{equation}\label{Xa-CT}
{(\bar{X}_H)}^A_a = -\frac{1}{4\theta}\, \underaccent{\tilde}{\varepsilon}_{ab}\, r_c \tilde{B}^{cb} + \nabla_a\phi\, \vert_{\partial\Sigma}\, ,
\end{equation}
which is the same as the value of the multiplier $\mu_a$ in the Maxwell + Chern-Simons theory, given in Eq.~(\ref{value_mu}).
As before, the continuity of ${(\bar{X}_H)}^A_a$ implies that its pullback to the boundary is ${(\bar{X}_H)}^{A}_a\vert_{\partial\Sigma}$,
and we obtain the same boundary condition (\ref{BCond3}), as in the Maxwell + Chern-Simons case. This condition can be rewritten in terms of $(\phi ,\tilde{\Pi}_\phi ; A_a, \tilde{\Pi}^a)$ and as a result we obtain the boundary condition given by Eq.~(\ref{BCond1}) for the Maxwell + Pontryagin theory. 

Let us summarize the situation of this third approach to the physical system. We started with the Maxwell + Pontryagin theory, that is defined by bulk terms alone, and through a canonical transformation generated by the Chern-Simons functional on $\Sigma$, the new canonical variables are those of the Maxwell theory. The new element is that the symplectic structure acquires a boundary contribution that corresponds to
that of the Chern-Simons theory on the boundary. This forces one to consider $A_a^\partial$, the pullback of the connection $A_a$ to the boundary $\partial\Sigma$, as new ``independent" boundary DOF (even when strictly speaking, they are completely determined by the bulk fields), in terms of computing partial derivatives. What we saw is that even when the Hamiltonian had only bulk terms (the Maxwell + Pontryagin), we {\it do} recover the bulk {\it and} boundary equations of motion together with the boundary conditions we found in Sec.~\ref{Sec:4}. Thus we have a third,``intermediate", description of the physical system.

Finally, we must point out that one could start with the Pontryagin contribution to the Hamiltonian, given by Eq.~(\ref{H_Pontryagin}), and through an integration by parts, rewrite it as a pure boundary term given by Eq.~(\ref{HamCCS}) (the bulk part vanishes due to the Bianchi identity). In this way we arrive at the Maxwell + Chern-Simons description in a very explicit fashion, since we recover the symplectic structure, Hamiltonian and boundary conditions of that theory. In that way, the equivalence of all descriptions becomes clear and manifest.

\section{Discussion}
\label{Sec:6}

The purpose of the manuscript is two-fold. First, we have put forward and extension of the standard Dirac-Regge-Teitelboim formalism for gauge field theories in regions with boundaries. This extension is needed when the dynamical structure of the theory demands a contribution to the symplectic structure from the boundary. This changes the standard DRT rules for differentiability, since now one is forced to consider
boundary contributions to the variations of functionals and  Hamiltonian vector fields. Consistency of the formalism requires to distinguish between bulk and boundary degrees of freedom when taking variations, and particular care in treating boundary and consistency conditions within Dirac's algorithm.

The second goal of this manuscript was to illustrate the general formalism with a concrete example that is 
particularly well suited for the task. In particular, it has the feature that it admits two descriptions, 
one in which one has a pure bulk theory and the second with a mixture of a bulk and a boundary term. The 
pure bulk theory is given by the Maxwell+Pontryagin action. In this case, since there is
no boundary contribution to the symplectic structure, one expects the DRT formalism to provide a 
consistent description. This is indeed the case. Thus, the Hamiltonian analysis is performed following 
Dirac's algorithm.  The only subtle point in the process is to properly handle the issue of boundary 
conditions. In order to have ``well defined" functionals we need to impose some boundary conditions, 
that allow the analysis of the dynamics of the system and the consistency conditions of the constraints. 
The boundary conditions are imposed on the space-like boundary $\partial\Sigma$, but the phase space is 
restricted to configurations that fulfil this conditions on the whole time-like boundary 
$[t_1,t_2]\times \partial\Sigma$. 

We have also performed the Hamiltonian analysis based on the geometric approach, where one of the main tasks is finding the Hamiltonian vector field that 
defines the evolution of the system. Here the main object to consider is the symplectic structure, that in this case, does not have boundary terms. In the process, 
the same boundary conditions as in the previous approach are obtained. An important point in both
approaches to the system, Dirac and GNH, is to regard all consistency conditions on the boundary as
boundary conditions, and not as secondary constraints. We shall elaborate more on this point below. 
%, but if one treats them as boundary constraints, the new ''tower`` of boundary constraints is obtained as well as some boundary conditions on the multipliers. All these constraints further restrict the phase-space variables on the boundary, that on the other hand restrict the Hamiltonian vector field on the boundary.

In the case of the Maxwell + Chern-Simons theory, the standard Dirac's analysis cannot be applied due to 
the presence of time derivatives in the boundary component of the canonical action. In this case canonical 
momenta can not be defined as local, differentiable functions, but have to be defined in the more general 
context as linear functionals. The previously accepted concepts of ``well defined" functionals and Poisson 
brackets, standard in the RT formalism, have to be extended. The corresponding symplectic structure has an 
additional boundary term, and the Hamiltonian vector field has a  boundary component as well. The 
components of the Hamiltonan vector field in the bulk are the same as in M+P theory and, therefore, the 
equations of the motion in the bulk coincide.
We also found that, apart from the primary constraint that also appears in the M+P theory, there are also 
boundary primary constraints, that turn out to be second class. Consistency conditions on the boundary 
variables yield boundary conditions relating bulk and boundary DOF, that coincide precisely with those 
found for Maxwell + Pontryagin. In this way we find complete agreement between the two descriptions but 
only if we treat the Maxwell + Chern-Simons theory with our new set of ``rules". Of particular relevance 
is the treatment of boundary conditions and boundary constraints in this case. It is important to properly 
distinguish them in order to have a consistent description. Failure to do so has caused some confusion in 
the literature, an issue that we hope to have clarified.

It is well known that The Maxwell + Pontryagin theory can be related to the pure Maxwell theory on the 
bulk. We follow this path and find yet a third description 
of our system. Starting with the M+P theory, we perform a canonical transformation defined by a generating 
functional (that turns out to be the Chern-Simons functional), and arrive to a theory that has only bulk 
constraints and Hamiltonian, but that in the process has acquired a boundary contribution to the 
symplectic structure, thus putting the system in  the category of those that need the ``extended" RT 
treatment. Therefore, we find that even when the Hamiltonian and constraints only have bulk contributions, 
we do have boundary contributions to the variations of various functionals and to the Hamiltonian vector 
fields. We recover then the description that we had previously obtained for the Maxwell + Chern-Simons.

The natural question is whether the formalism here proposed in generic enough to deal with other physical 
systems of interest. We have also analysed in detail the case of (weakly) isolated horizons that are a 
useful way of modelling black holes in equilibrium via boundary conditions imposed on an internal boundary 
of spacetime. As mentioned in the introduction, in early papers on the subject \cite{IH-old}, some of the 
ideas for considering boundary terms were introduced, but without having a complete and consistent 
formulation as here presented. In a separate paper, that can be considered as a follow-up to the 
present manuscript, we have analysed a more general case of isolated horizons than in \cite{IH-old}, and 
have performed a careful canonical analysis of the system. That manuscript shall be published soon 
\cite{CRV-Canonical}.

\begin{acknowledgments}
This work was in part supported by DGAPA-UNAM IN114620 and CONACyT 0177840 grant,
and by CIC, UMSNH. T. Vuka\v sinac would like to thank CCM-UNAM, for its hospitality during her sabbatical year. We thank I. Rubalcava for discussions in the early part of this work.
\end{acknowledgments}

\appendix
\section{Some useful results}
\label{Sec:A}
In this Appendix we shall collect some useful formulae for integration on manifolds, when we have both the Levi Civita symbol, that requieres no underlying metric, and the more common case of a volume integral defined by the metric. This is a review of known results, but might be useful for the reader since we are using both integration methods in our model.

\subsection{Levi-Civita symbols and densities}
\label{Sec:A.1}

Here we will present with more details some results that we used throughout Sections \ref{Sec:3}, \ref{Sec:4} and \ref{Sec:5}. For details see, for example, \cite{Poisson} and \cite{Wald}.

Let us start with 4-dim. Levi-Civita symbol that is defined, when we think of indices as taking numerical values (and for a moment, departing from the abstract index notation), as

\begin{equation}
    \underaccent{\tilde}{\varepsilon}_{abcd}=
    \begin{cases}
      \ \ \, 1, & \text{if}\ abcd \ \text{is an even permutation of}\ 0123 \\
      -1, & \text{if}\ abcd \ \text{is an odd permutation of}\ 0123  \\
      \ \ \, 0, & \text{if any two indices are equal}\, .
    \end{cases}
  \end{equation}
By definition 
\be
\tilde{\varepsilon}^{abcd} := \underaccent{\tilde}{\varepsilon}_{abcd}\, .
\ee
The corresponding Levi-Civita tensors are defined as
\begin{eqnarray}
{\varepsilon}_{abcd} &:=& \sqrt{-g}\,  \underaccent{\tilde}{\varepsilon}_{abcd}\, ,\\
{\varepsilon}^{abcd} &:=& -\frac{1}{\sqrt{-g}}\, \tilde{\varepsilon}^{abcd}\, ,
\end{eqnarray}
where $g$ is the determinant of the Lorentzian metric $g_{ab}$ on 4-dim. spacetime $\mathcal{M}$. 

Suppose now that we have a foliation $M=I\times\Sigma$, where $M\subset \mathcal{M}$. 
Then \cite{Wald},
\be
\varepsilon^{abcd}=-4n^{[a}\varepsilon^{bcd]}\, ,
\ee
where $n^a$ is an exterior unit normal to $\Sigma$ and $\tilde{\varepsilon}^{bcd}$ is a 3-dim. (spatial) Levi-Civita tensor on $\Sigma$. 
The foliation of spacetime induces the decomposition for the spacetime metric $g_{ab}$ into the six independent components of the Euclidean spatial metric $h_{ab}$, the lapse function $N$ and the  shift vector $N^a$. Lapse and shift being respectively the normal and tangential components of the evolution vector field $t^a=Nn^a+N^a$, where again, $n^a$ denotes the future directed normal to the foliation and $N=-\left|(\md t)_\mu(\md t)^\mu\right|^{-1/2}$. In coordinates $(t,y^a)$ adapted to the foliation the line element reads
\begin{equation} \label{ADMmetric}
g_{ab}\md x^a \md x^b=(-N^2+h_{ab}N^aN^b)\md t^2+2h_{ab}N^b\md t\,\md y^a + h_{ab}\md y^a \,\md y^b\,.
\end{equation}
It follows that
\be
- \frac{1}{\sqrt{-g}}\, \tilde{\varepsilon}^{abcd}= -\frac{4}{\sqrt{h}}\, n^{[a}\tilde{\varepsilon}^{bcd]}\, 
\ee
where $h$ is the determinant of the metric induced on $\Sigma$, $\tilde{\varepsilon}^{bcd}$ is a 3-dim. Levy-Civita symbol (a tensor density of weight one on $\Sigma$) and the minus sign on the RHS follows from the fact that $n^a$ is time-like, $n_an^a=-1$. Since, $\sqrt{-g}=N\sqrt{h}$ and $Nn^a = t^a - N^a$, we obtain \cite{romano}
\be\label{A8}
\tilde{\varepsilon}^{abcd}= 4t^{[a}\tilde{\varepsilon}^{bcd]}\, .
\ee
We used that $N^{[a}\tilde{\varepsilon}^{bcd]}=0$ since all the indices are spatial.

Similarly, for $\mathcal{D}=I\times\partial\Sigma$ we have
\be\label{A9}
\tilde{\varepsilon}^{abc}= 3t^{[a}\tilde{\varepsilon}^{bc]}\, ,
\ee
where $\tilde{\varepsilon}^{bc}$ is a 2-dim. Levy-Civita symbol (tensor density of weight one on $\partial\Sigma$). 

\subsection{Integration by parts}
\label{Sec:A.2}

The Stokes' theorem for vector density of weight one on $M$, $\tilde{B}^a$, takes the form
\be\label{A10}
\int_{M}\md^4x\, \nabla_a\tilde{B}^a = \int_{\partial M}\md S_a\, \frac{\tilde{B}^a}{\sqrt{-g}}\, ,
\ee
where $\md S_a$ is a directed hyper-surface element on $\partial M$,
 \be
\md S_a = \epsilon\, r_a\sqrt{|h|}\, \md^3 y\, ,
\ee
where $\epsilon =r_ar^a$. It can be rewritten as
\be\label{Stokes}
\int_{M}\md^4x\, \nabla_a\tilde{B}^a = \int_{\partial M}\md^2y\, r_a\tilde{B}^a\, ,
\ee
where on the RHS we use the same notation $\tilde{B}^a$ for a vector density of weight one on $\partial\Sigma$.  Namely, 
\be
\frac{(\tilde{B}^a)_{\rm LHS}}{\sqrt{-g}}=\frac{(\tilde{B}^a)_{\rm RHS}}{\sqrt{|h|}}\, .
\ee
The  Pontryagin term can be written as 
\be
S_{\rm P}= -\frac{\theta}{4}\int_M \md^4x\, \tilde{\varepsilon}^{abcd}F_{ab}F_{cd}= 
-\frac{\theta}{2}\int_{M} \md^4 x\, \nabla_a(\tilde{\varepsilon}^{abcd}A_b F_{cd})\, .
\ee
Since $M=\Sigma_1\cup\Sigma_2\cup\mathcal{D}$, and the variations of fields on $\Sigma_1$ and $\Sigma_2$ vanish, $S_{\rm P}$ is equivalent to the Chern-Simons term on $\mathcal{D}=I\times\partial\Sigma$. Using (\ref{A10}), and ignoring the contributions from the initial and final Cauchy surfaces, we obtain that
\be\label{A15}
S_{\rm P}=S_{\rm CS}=\frac{\theta}{2}\int_{\mathcal{D}} \md S_a\, \varepsilon^{abcd}A_b F_{cd}=
\frac{\theta}{2}\int_{\mathcal{D}} \md^3 y\, \tilde{\varepsilon}^{bcd}A_b F_{cd}\, ,
\ee
since $\md S_a =r_a\sqrt{|q|}\,\md^3y$, where $r^a$ is an exterior (spatial) unit normal to $\mathcal{D}$ and $q_{ab}$ is the induced metric on $\partial\Sigma$, of the signature $(-1,1,1)$. In this case we have used the following relations
\be
\varepsilon^{abcd}=-4r^{[a}\varepsilon^{bcd]}=\frac{4}{\sqrt{|q|}}\, r^{[a}\tilde{\varepsilon}^{bcd]}
\ \ \ \Rightarrow\ \ \ \md S_a\,\varepsilon^{abcd}=\md^3 y\,\tilde{\varepsilon}^{bcd}\, .
\ee

\end{document}